\documentclass[twocolumn]{aastex62}

\shorttitle{Recombining Plasma in W49B}
\shortauthors{Holland-Ashford et al.}

\newcommand{\ltsima}{$\; \buildrel < \over \sim \;$}
\newcommand{\simlt}{\lower.5ex\hbox{\ltsima}}

\def\arcdeg{\hbox{$^\circ$}}
\def\arcmin{\hbox{$^\prime$}}
\def\arcsec{\hbox{$^{\prime\prime}$}}
\usepackage{hyperref}
\usepackage{xcolor}

\newcommand{\err}[2]{$^{+#1}_{-#2}$}

\begin{document}

\title{Spatially-Resolved Study of Recombining Plasma in W49B Using {\it XMM-Newton}}

\author{Tyler Holland-Ashford}
\affil{Department of Astronomy, The Ohio State University, 140 W. 18th Ave., Columbus, OH 43210, USA}
\affil{Center for Cosmology and AstroParticle Physics, The Ohio State University, 191 W. Woodruff Ave., Columbus, OH 43210, USA}

\author{Laura A. Lopez} 
\affil{Department of Astronomy, The Ohio State University, 140 W. 18th Ave., Columbus, OH 43210, USA}
\affil{Center for Cosmology and AstroParticle Physics, The Ohio State University, 191 W. Woodruff Ave., Columbus, OH 43210, USA}
\affil{Niels Bohr Institute, University of Copenhagen, Blegdamsvej 17, 2100 Copenhagen, Denmark}

\author{Katie Auchettl}
\affil{School of Physics, The University of Melbourne, Parkville, VIC 3010, Australia}
\affil{ARC Centre of Excellence for All Sky Astrophysics in 3 Dimensions (ASTRO 3D)}
\affil{DARK, Niels Bohr Institute, University of Copenhagen, Lyngbyvej 2, 2100 Copenhagen, Denmark}
\affil{Department of Astronomy and Astrophysics, University of California, Santa Cruz, CA 95064, USA}

\begin{abstract}

W49B is the youngest SNR to date that exhibits recombining plasma. The two prevailing theories of this overionization are rapid cooling via adiabatic expansion or through thermal conduction with an adjacent cooler medium. To constrain the origin of the recombining plasma in W49B, we perform a spatially-resolved spectroscopic study of deep {\it XMM-Newton} data across 46 regions. We adopt a 3-component model (with one ISM and two ejecta components), and we find that recombining plasma is present throughout the entire SNR, with increasing overionzation from east to west. The latter result is consistent with previous studies, and we attribute the overionization in the west to adiabatic expansion. However, our findings contrast these prior works as we find evidence of overionization in the east as well. As the SNR is interacting with molecular material there, we investigate the plausibility of thermal conduction as the origin of the rapid cooling. We show that based on the estimated timescales, it is possible that small-scale thermal conduction through evaporation of clumpy, dense clouds with a scale of 0.1--1.0~pc can  explain the observed overionization in the east. 

\end{abstract}


\section{Introduction} \label{sec:intro}

X-ray studies have revealed that many supernova remnants (SNRs) have recombining plasmas (RPs; e.g., \citealt{ozawa09,yamaguchi09,ohnishi11,uchida12,lopez13b,2014PASJ...66..124S, 2015ApJ...810...43A,2015ApJ...808...77U, washino16,2017ApJ...847..121A, matsumura17,okon18,suzuki18}). In particular, their spectra show radiative recombination continua (RRCs) which arise when electrons collide and recombine with ions. RPs are overionized, where the electron temperature $kT_{\rm e}$ from the continuum is lower than the ionization temperature $kT_{\rm z}$ given by line ratios \citep{kawasaki02,kawasaki05}, in contrast to a collisional ionization equilibrium (CIE) plasma where these temperatures are equal. Most young SNRs (with ages $\lesssim10^{3}$~years) are underionized, whereas older SNRs have reached CIE. However, in SNRs where rapid cooling has occurred, overionization is possible because the  recombination timescale is longer than the cooling time (see the review by \citealt{yamaguchi20}).

The origin of RPs is debated: the two proposed scenarios are rapid cooling arising from adiabatic expansion or from thermal conduction (the timescale for cooling via radiation is much too long; see \citealt{masai94} and \citealt{kawasaki02}). After gas is shock-heated to high temperatures, rapid cooling via adiabatic expansion occurs when the gas breaks through to a low-density interstellar medium (ISM; e.g., \citealt{moriya12,shimizu12}). The electrons cool more rapidly than the ions, producing an overionized plasma (see e.g., \citealt{itoh89,yamaguchi09}). Thermal conduction, on the other hand, occurs when hot ejecta cools by the exchange of heat with cooler material (e.g., \citealt{cox99,shelton99}) that can have a short enough timescale to produce an overionized plasma \citep{kawasaki02}. Among the SNRs with evidence of RP, all are interacting with molecular clouds, indicating that this interaction is tied somehow to the rapidly cooling plasma. All SNRs with RPs are of the mixed-morphology class \citep{rho98}, which tend to be mature SNRs with ages of $\sim$4000--20000~years (see Section 10.3 of \citealt{vink12}).

W49B (G43.3$-$0.2) is the youngest SNR with overionized plasma ($\sim$1000--4000~years: \citealt{pye84,hwang00}). Thus, W49B's RP demonstrates that cooling can occur even in the early stages of SNR evolution. W49B is interacting with a molecular cloud on its eastern side \citep{keohane07,zhu14}, and it is expanding into less-dense ISM on its western side. Consequently, W49B is an ideal target to explore the origin of RPs and overionization in SNRs.

W49B was one of the first SNRs where signatures of a recombining/overionized plasma were discovered. Using integrated {\it ASCA} data, \cite{kawasaki05} measured the intensity ratios of He-like to H-like lines of Ar and Ca, and they found $kT_{\rm z} \sim 2.5$~keV $\gtrsim kT_{\rm e} \sim 1.8$~keV. They proposed that thermal conduction was the origin of the overionization and found that the thermal conduction timescale was less than the recombination timescale. 

Subsequently, evidence of a recombining plasma in W49B was identified by \cite{ozawa09} using {\it Suzaku} observations: they found an RRC feature from He-like Fe at 8.830~keV and derived $kT_{\rm z} \sim 2.7$~keV. \cite{miceli10} localized spatially the recombining plasma using {\it XMM-Newton} data by mapping the hardness ratio of the count rate in the 8.3--12~keV band to that in the 4.4--6.2~keV band. They found that the hardness ratio is enhanced in the center and west of W49B, which these authors attributed to the Fe RRC. However, the limited counts in these {\it XMM-Newton} observations precluded a spatially-resolved spectral modeling of these features, and the enhanced hardness ratio could have arisen from other emission mechanisms (e.g., bremsstrahlung, synchrotron) or from high background. 

\cite{lopez13a} presented a deep, 220-ks {\it Chandra} observation of W49B, and \cite{lopez13b} analyzed this data to measure $kT_{\rm z}$ in 13 regions across the SNR using the He-like to H-like line ratios of S and Ar. They showed that the SNR is overionized in the west, where $kT_{\rm e} < kT_{\rm z}$, consistent with the adiabatic cooling scenario and the results of \cite{miceli10}. \cite{zhou18} performed a spatially-resolved spectroscopic study using the {\it Chandra} data, and they found increased ionization timescale $\tau = n_{\rm e} t$ and $kT_{\rm e}$ in the west, with lower recombination ages of $\sim$2000~years there compared to that in the east ($\sim$6000 years).

Using {\it NuSTAR} observations, \cite{yamaguchi18} mapped the ratio of the Fe RRC line (at 8.8--10~keV) to the Fe He-$\alpha$ line (at 6.4--6.8~keV) in W49B. They found that this ratio is enhanced in the west, implying greater overionization there. In addition, \cite{yamaguchi18} conducted a spatially-resolved spectral analysis of 12 regions across W49B, fitting data with a single RP model and linking the initial temperature $kT_{\rm init}$ across the regions. The western regions exhibited lower $\tau$ and $kT_{\rm e}$ than the eastern regions, again supporting the adiabatic cooling scenario. 

Recently, \cite{sun20} analyzed 2004 and 2014 {\it XMM-Newton} observations of W49B with a total effective exposure time of $\sim$230 ks. They performed a global spectral analysis and showed that one CIE $+$ two RP components fit the spectra well, with $kT_{\rm init}$ values of $\sim$2.4 and 4.5~keV. These results are consistent with the spatially-resolved {\it Chandra} study of W49B performed by \cite{zhou18}, who found initial temperatures near $\sim$2.5 keV in the east and $\gtrsim$5keV in the west. \cite{sun20} suggested the high $kT_{\rm init}$ component might have arisen from either thermal conduction or adiabatic cooling, whereas the lower $kT_{\rm init}$ component likely originated from adiabatic cooling. Hydrodynamical simulations by \cite{zhou11} and \cite{zhang19} supported the possibility that both cooling scenarios together could explain the RP in W49B. 

In this paper, we use the deep {\it XMM-Newton} observations of W49B to map the RP in order to ascertain the origin of the rapid cooling in the SNR. This work improves upon that of \cite{sun20} by performing a spatially-resolving analysis of the deep {\it XMM-Newton} data, and we make use of XSPEC's RP models which were unavailable at the time of \cite{lopez13b}. The analysis approach is similar to \cite{zhou18} and \cite{yamaguchi18}, taking advantage of {\it XMM-Newton}'s better hard X-ray response than {\it Chandra} and improved spectral resolution relative to {\it NuSTAR}.

\begin{figure}[t!]
\begin{center}
\includegraphics[width=\columnwidth]{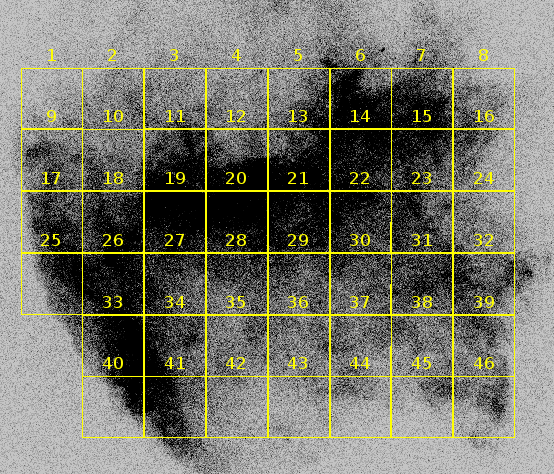}
\end{center}
\caption{A {\it Chandra} image \citep{lopez13b} of W49B with the analyzed regions overplotted. Note that the region numbers are indicated by the values {\it above} the boxes. North is up, and East is left.} 
\label{fig:regions}
\end{figure}

\begin{figure*}[t!]
\begin{center}
\includegraphics[width=\textwidth]{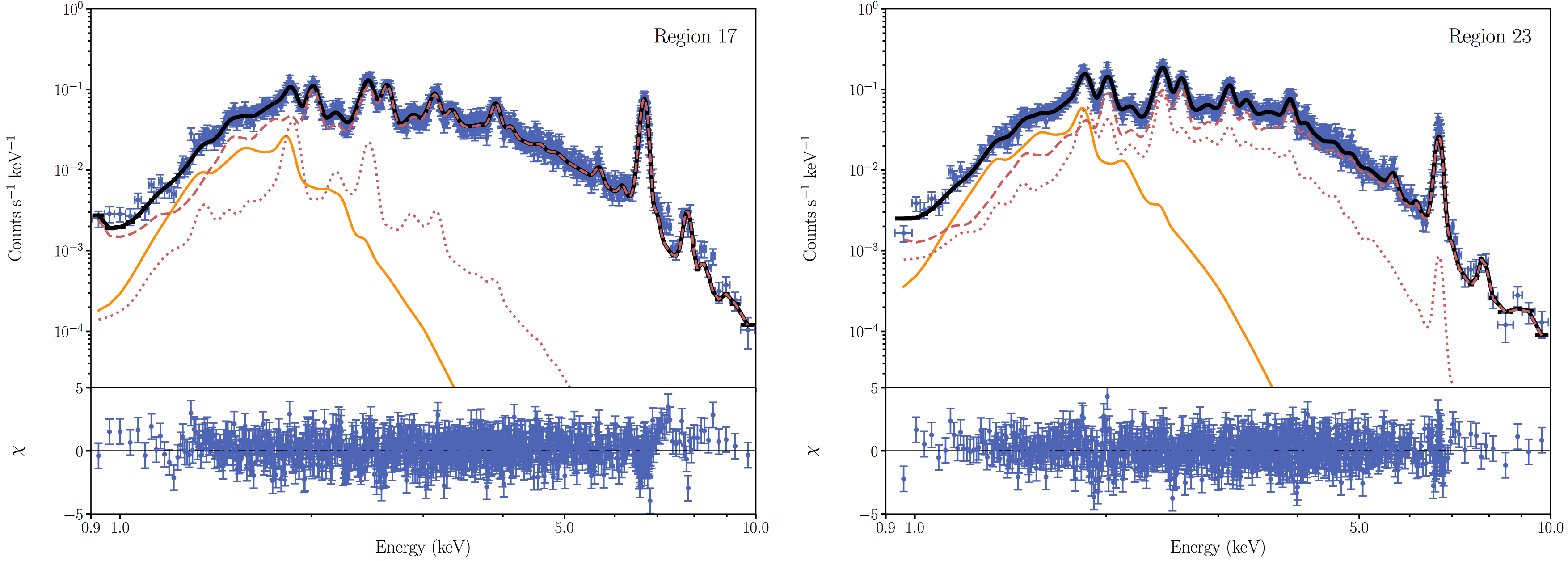}
\end{center}
\vspace{-5mm}
\caption{Example spectra and fits from region 17 (left) and region 23 (right), where the cooler ejecta component is in CIE or is overionized, respectively. The black line represents the best-fit model, the orange line is the ISM component, and the red dotted (dashed) line is the cooler (hotter) ejecta components.} 
\label{fig:spectra}
\end{figure*}

The paper is organized as follows. In Section~\ref{sec:analysis}, we present the observations as well as our methods of data reduction and spectral fitting. In Section~\ref{sec:results}, we describe the results and show maps of the best-fit parameters, discussing how they relate to the presence and origin of the overionized plasma in W49B. Finally, in Section~\ref{sec:conclusions}, we summarize our findings and the implications.

\section{Observations and Data Analysis}\label{sec:analysis}
\subsection{{\it XMM-Newton} Data Reduction}

W49B has been observed using the \textit{XMM-Newton} Observatory five times, with three observations in 2004 (ObsIDs 0084100401--0084100601; PI: Decourchelle) for 39.3~ks and two observations in 2014 (ObsIDs 0724270101 and 0724270201; PI: Lopez) for 189.7~ks. We only consider the recent pair of observations in order to limit the systematic effects from the instruments' spectral response. These observation were taken on 2014 April 17--19, with both the MOS and PN detectors operated in the full-frame mode with the medium filter, and have exposure times of 118.5~ks and 71.2~ks, respectively. The SNR (with a diameter of $\sim$4\arcmin: \citealt{green19}) is fully enclosed by the field of view of the detectors. 

To process the data, we used the {\it XMM-Newton} Science System (SAS) version 15.0.0 \citep{gabriel04} and the most up-to-date calibration files to produce the data products for our analysis. As \textit{XMM-Newton} suffers from both proton flares and a high background, we made count rate histograms using events with an energy between 10--12 keV and removed the time intervals which were contaminated by a high background or flares to produce our cleaned events. We found that these observations are only partially affected by high background/proton flares, giving an effective exposure time of 141.4 ks, 150.3 ks, and 90 ks for the MOS1, MOS2 and PN detectors, respectively. For all of the detectors, we used the standard screening procedures and screening set of FLAGS as suggested in the current SAS analysis threads\footnote{\url{https://www.cosmos.esa.int/web/xmm-newton/sas-threads}} and \textit{XMM-Newton} Users Handbook\footnote{\url{https://xmm-tools.cosmos.esa.int/external/xmm_user_support/documentation/uhb/}}. As W49B is located in the Galactic plane, it is possible that both Galactic Ridge X-ray emission and the Cosmic X-ray Background can contribute non-negligibly to the observed emission. To correct for this contamination, we used \textsc{evigweight} on all of the cleaned event files which allows us to take vignetting into account. All analysis products and results presented below are from these cleaned, filtered and vignetting-corrected event files.

We extracted spectra from 46 0.5\arcmin$\times$0.5\arcmin\ regions covering the spatial extent of the SNR (see Figure \ref{fig:regions}) using the SAS task \textsc{evselect}. Spectral response and effective area files for each detector were produced using the SAS tasks \textsc{arfgen} and \textsc{rmfgen}, respectively. Given the high signal from W49B, we accounted for the background using background subtraction from a 136\arcsec$\times$58\arcsec\ rectangular background region centered at ($\alpha$,$\delta$)=($19
^{h}11^{m}18.1^{s},-9^{\circ}11\arcmin26.3\arcsec$). We combined the MOS1 and MOS2 spectra for each observation using the HEASARC command \textsc{addascaspec}.

\subsection{Spectral Fitting} \label{subsec:fitting}

The spectral fitting was performed using the X-ray analysis software XSPEC Version 12.9.0 \citep{arnaud96} and ATOMDB Version 3.0.9 \citep{smith01,foster12} over an energy range of $0.9-10.0$~keV. Each spectrum was grouped with a minimum of 25 counts per energy bin using the FTOOLS command \textsc{grppha} and fitted using $\chi^2$ statistics. We adopted solar abundances from \cite{aspl09}.

We fit the spectra of each region using an absorbed model with one ISM component and two ejecta components. In every region, we found that a single ejecta component was insufficient to adequately fit the data (e.g., producing large residuals around prominent emission lines) and that it was necessary to include at least one RP component. We attempted using an underionized, overionized, or CIE model to describe the lower-temperature ejecta component, and we found that either an overionized or CIE component produced the best fit, depending on the region analyzed. We initially fit each region with two RP components, and if the ionization timescale $\tau$ exceeded 3$\times10^{12}$~cm$^{-3}$~s, we fit the lower-temperature ejecta with a CIE component, since 90\% of the material at temperatures of a few $\times 10^7$ K reach CIE by $\tau$ of a few $\times10^{12}$ cm$^{-3}$~s \citep{smith10}. In these cases, using a CIE model to fit the cooler ejecta resulted in a lower reduced $\chi^2$ than using a RP component.

Our final model for each region was either \\ \texttt{phabs}$\times$(\texttt{vapec}+\texttt{vvapec}+\texttt{vvrnei}) or \\ \texttt{phabs}$\times$(\texttt{vapec}+\texttt{vvrnei}+\texttt{vvrnei}). The first \texttt{vapec} component, a model for fitting a plasma in CIE with non-solar abundances, represents the shock-heated ISM/CSM. For that component, we fixed Mg to 0.3~$Z_{\sun}$ to match the fits of \citealt{sun20}, and other abundances were set to 1~$Z_{\sun}$. The middle \texttt{vvapec} or \texttt{vvrnei} component represents the lower-temperature ejecta that either has reached CIE or is overionized, respectively. The final \texttt{vvrnei} component represents the hot overionized ejecta. As a \texttt{vvrnei} model reflects overionized plasma, in addition to the typical fitting parameters (e.g., current electron temperature $kT_e$, ionization timescale $\tau$, normalization $\texttt{norm}$, and abundances of various elements), the model also includes the parameter $kT_{\rm{init}}$. This parameter captures the initial temperature of the plasma before rapid cooling occurred. In order to constrain the fits, we found it necessary to link $kT_{\rm init}$ of the two RP ejecta components.

Additionally, we tied the abundances between the two ejecta components and allowed Mg, Si, S, Ar, Ca, Cr, Mn, Fe, and Ni to vary. Due to the high column density $N_{\rm H}$ and a corresponding lack of line detections below $\sim$1.1~keV, we froze all elements lighter than Mg to solar metallicity. Generally, we found that the emission below $\sim$1.9~keV is largely produced by the ISM and cooler ejecta component and is attenuated by the high $N_{\rm H}$ toward W49B. As such, we note that there may be a degeneracy between the fit parameters of these components (e.g., N$_{\rm H}$, kT$_{\rm ISM}$, kT$_{\rm e,1}$, $\tau_{1}$, and the Si abundance). Above $\sim$1.9~keV, the hotter ejecta component dominates the flux, enabling more robust measurement of the associated parameters (e.g., $kT_{e,2}$, $kT_{\rm init}$, $\tau_{2}$, and the S, Ar, and Ca abundances) to assess overionization.

\begin{figure*}
\begin{center}
\includegraphics[width=\textwidth]{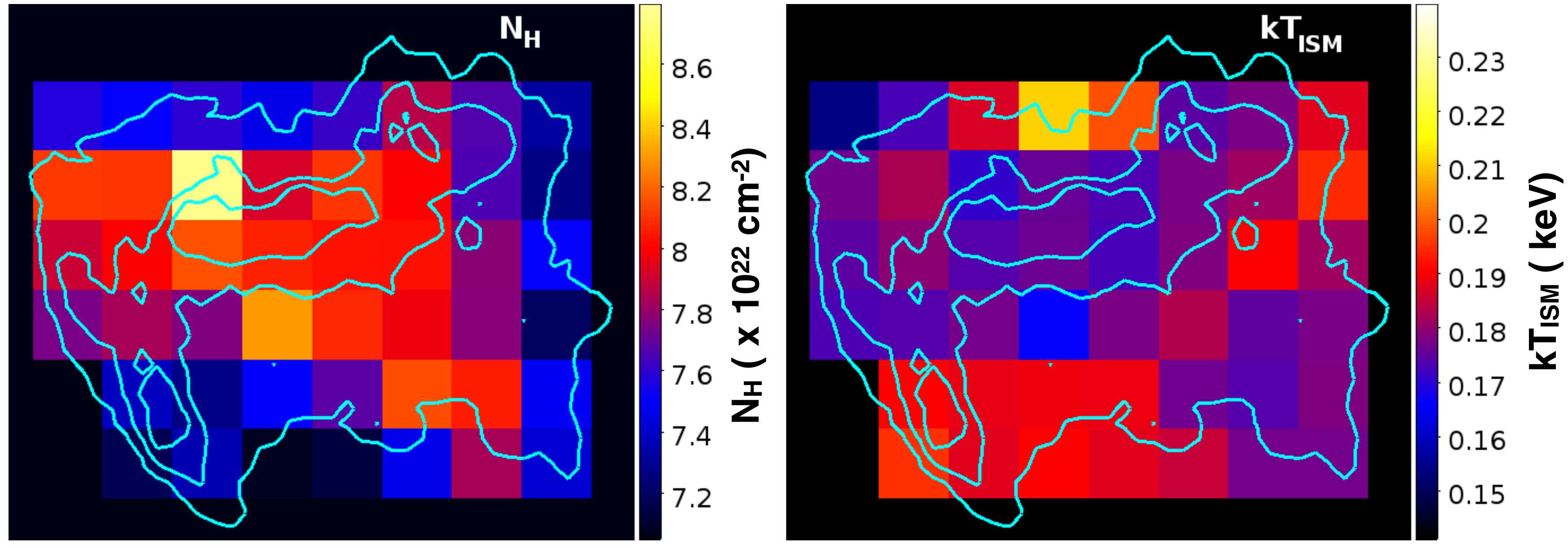}
\caption{Maps of best-fit $N_{\rm{H}}$ (left panel) and $kT_{\rm{ISM}}$ (right right) values. Cyan contours of the $0.5-7.0$~keV emission (from {\it Chandra}; \citealt{lopez13a}) are overplotted. North is up, and East is left.}
\vspace{3mm}
\label{fig:firstparams}
\end{center}
\end{figure*}

\subsubsection{Ionization Temperature and Fe RRC Measurements} \label{subsec:iontempcalc}

Past studies of overionization in W49B (e.g., \citealt{kawasaki05, lopez13b, sun20}) have calculated flux ratios between elements' (Si, S, Ar, Ca) Ly-$\alpha$ and He-$\alpha$ emission lines in order to estimate the ionization temperatures $kT_{\rm z}$ of each element. These studies have shown that higher-mass elements have greater $kT_{\rm z}$ (i.e., $kT_{\rm z, Ar}\approx$ 2~keV; $kT_{\rm  z, Ca}\approx$ 2.5~keV) than lower-mass elements (i.e., kT$_{\rm z, Si}\approx$ 1.1~keV; kT$_{\rm z, S}\approx$ 1.6~keV) in W49B. \cite{lopez13b} concluded that the origin of these differences is unlikely to be physical, since heavier elements require longer ionization timescales to reach CIE \citep{smith10}. However, \cite{sun20} suggested that the differences may arise because the Si and S are predominantly associated with the cooler RP, whereas the Ar and Ca are from the higher-temperature RP.

\begin{figure*}
\begin{center}
\includegraphics[width=\textwidth]{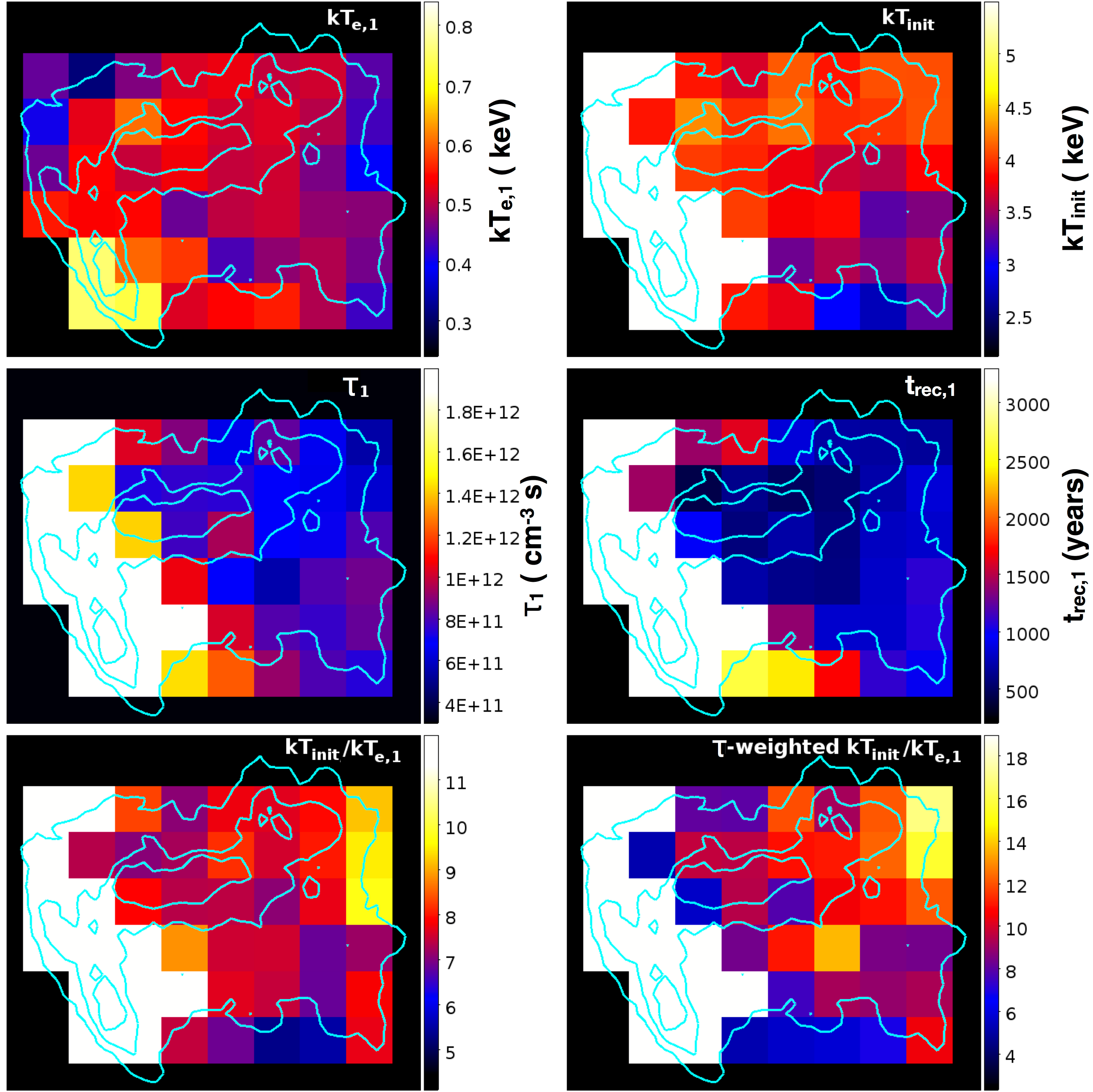}
\caption{Maps of the best-fit parameters of the lower-temperature (cooler) ejecta component with cyan contours overplotted from the broad-band X-ray emission (from {\it Chandra}; \citealt{lopez13b}). The white squares in the maps reflect the plasma in CIE. The $t_{\rm rec}$ is estimated by dividing $\tau_{1}$ by the estimated electron density $n_{\rm e,1}$. North is up, and East is left.}
\vspace{3mm}
\label{fig:ej1params}
\end{center}
\end{figure*}

We examine the spatial distribution of $kT_{\rm z}$ throughout the SNR to investigate the possible origin of these differing temperatures. To do this, we measure the continuum-subtracted fluxes of the Ly-$\alpha$ and He-$\alpha$ lines for Si, S, Ar, and Ca. We account for the thermal continuum using the AtomDB NoLine\footnote{http://www.atomdb.org/noline.php} model \textsc{apecnoline}, an XSPEC model that includes no emission lines. We adopt the best-fit values of our full spectral analysis (e.g., $N_{\rm H}$, $kT_{\rm e}$, norm, $\tau$) into a \texttt{phabs}*(\texttt{apec}+\texttt{apec}+\texttt{apec}) model, where the 3 \texttt{apec} (noline) models represent the ISM, cool ejecta, and hot ejecta components. From each ejecta \texttt{vvrnei} (or \texttt{vvapec}) component, we subtract our derived \textsc{apecnoline} continuum to obtain the continuum-subtracted line flux of each element. We perform this calculation for the He-$\alpha$ and Ly-$\alpha$ line of Si, S, Ar, and Ca.  We then convert these flux ratios to $kT_{\rm z}$ using the relations shown in Figure~6a of \cite{sun20} and created from AtomDB data tables.

We adopt the same approach as above to measure and map the flux of the Fe RRC line (at $E_{\rm edge}$ = 8.83~keV) as well (see Section~\ref{subsec:FeRRC} for more details).

\section{Results} \label{sec:results}

Table~\ref{table:fits} in the Appendix lists the best-fit parameters for each of the 46 regions analyzed, including 1-$\sigma$ errors on all quantities that were derived using the XSPEC \texttt{error} command. Two example spectra and their best fits are presented in Figure~\ref{fig:spectra}. The left panel shows the best-fit spectra associated with region 17, where the cooler ejecta is found to be in CIE, while the right panel shows the spectra from region 23, where both ejecta components are overionized. In Sections~\ref{subsec:firstparams}--\ref{subsec:ej2_maps}, we consider the results associated with each spectral component and the implications.

\subsection{$N_{\rm H}$ and $kT_{\rm ISM}$ Maps} \label{subsec:firstparams}

In Figure~\ref{fig:firstparams}, we show the best-fit N$_{\rm{H}}$ and kT$_{\rm{ISM}}$ maps. We find a range in best-fit $N_{\rm H}$ of (7.2--8.5)$\times10^{22}$~cm$^{-2}$ and best-fit $kT_{\rm ISM}$ of 0.16--0.20~keV. These values are consistent with previous measurements by \cite{lopez13b}, \cite{zhou18}, and \cite{sun20}.

$N_{\rm H}$ is elevated in the center, east, and southwest, whereas it is comparatively lower in the southeast, northeast, and exterior of W49B. These results suggest that the dense material in front of (or associated with) the SNR is not uniform. We note that CO and warm H$_{2}$ has been detected in the east and southwest of W49B \citep{lacey01,keohane07,zhu14}, consistent with the enhanced $N_{\rm H}$ in those directions. The elevated $N_{\rm H}$ in region~11 (of $N_{\rm H} = (8.76$\err{0.03}{0.07})$\times10^{22}$~cm$^{-2}$) corresponds to a ``hole" seen in {\it Chandra} images where X-rays below $\sim$2.5~keV are attenuated (see Figure~3 of \citealt{lopez13b}) which suggests the existence of dense foreground material there. 

Our $kT_{\rm ISM}$ values are clustered around $\approx$ 0.18~keV and are relatively constant across the SNR varying only by $\sim$20\% in the 46 regions. We note that our $kT_{\rm ISM}$ map tends to be anti-correlated with the best-fit $N_{\rm H}$ (where regions of high $N_{\rm H}$ have lower $kT_{\rm ISM}$). As discussed in Section~\ref{subsec:fitting}, the $N_{\rm H}$ and $kT_{\rm ISM}$ are determined largely from the flux below $\sim$2.5~keV, so it is possible that these components are partially degenerate with one another.

\subsection{Cooler Ejecta Component Maps}
\label{subsec:ej1_maps}

\begin{figure*}
\begin{center}
\includegraphics[width=\textwidth]{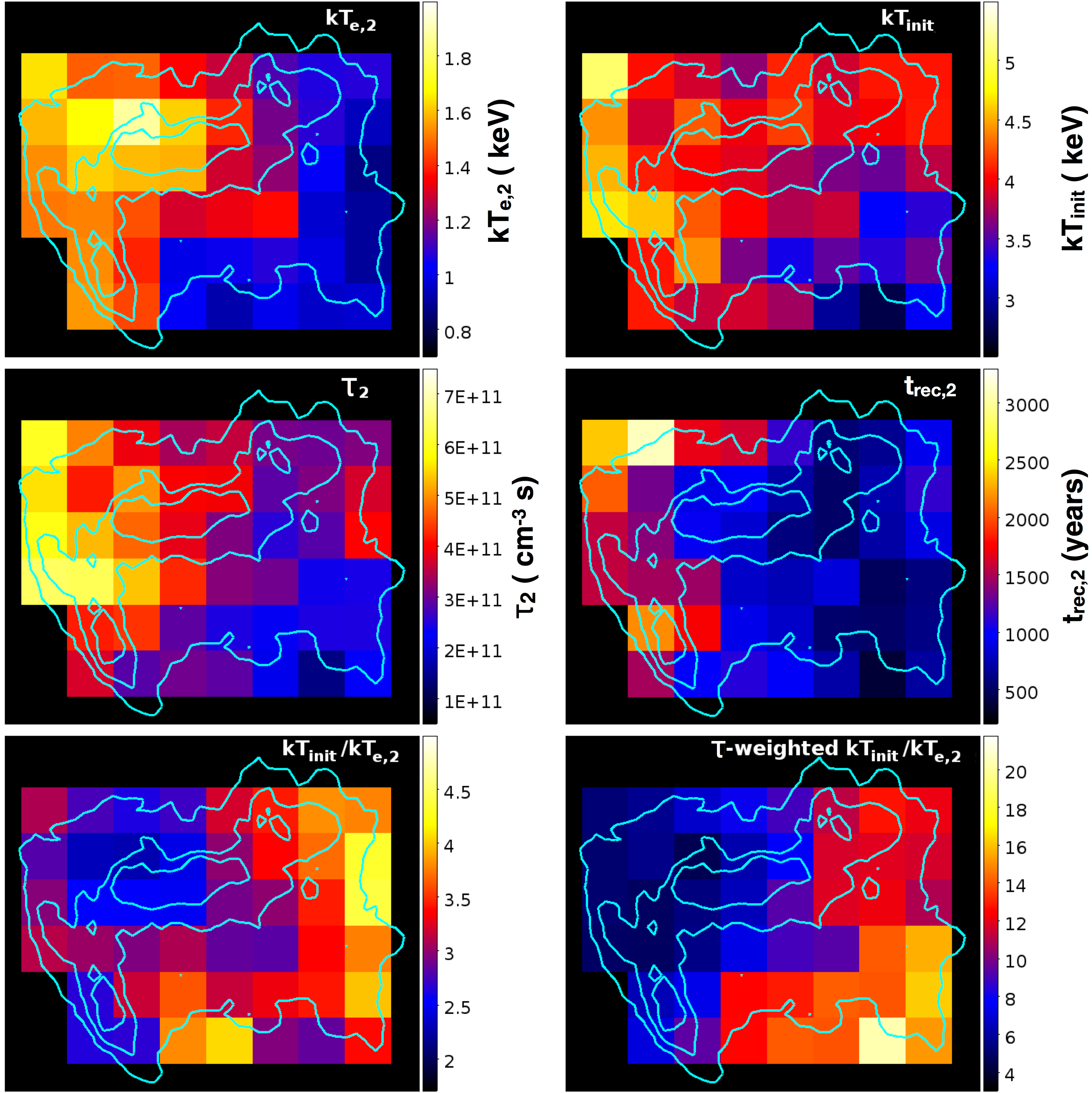}
\caption{Maps of the best-fit parameters of the hotter ejecta component with cyan contours overplotted from the broad-band X-ray emission (from {\it Chandra}; \citealt{lopez13b}). The $t_{\rm rec,2}$ is estimated by dividing $\tau_{2}$ by the estimated electron density $n_{\rm e,2}$. North is up, and East is left.}
\vspace{3mm}
\label{fig:ej2params}
\end{center}
\end{figure*}

In Figure~\ref{fig:ej1params}, we show the maps of the parameters associated with the cooler ejecta component: the current electron temperature $kT_{e,1}$, the initial temperature $kT_{\rm init}$, the ionization timescale $\tau_{1}$, the recombination timescale t$_{\rm rec,1}$, and the ratio of $kT_{e,1}/kT_{\rm init}$. We find a range in best-fit $kT_{e,1}$ of 0.31--0.76~keV and best-fit $kT_{\rm init}$ of 2.72--4.26 keV. Our results are consistent with those of \cite{sun20} who found $kT_{e,1} = 0.64\pm$0.01~keV and $kT_{\rm init}$ values of 2.42\err{0.06}{0.03}~keV and 4.54\err{0.17}{0.07}~keV of their cooler and hotter components, respectively, from the integrated {\it XMM-Newton} spectrum. Both of these temperature maps show enhanced temperatures along the bar, and $kT_{\rm e,1}$ is additionally enhanced in the southeast.

The best-fit ionization timescales $\tau_{1}$ of the overionized cooler plasma are $\sim$(7--14)$\times10^{11}$~cm$^{-3}$~s. We find that the cooler component has reached CIE in the east, as denoted by the white squares in the parameter maps in Figure~\ref{fig:ej1params}. By contrast, the best-fit $\tau_{1}$ values in the west suggest that the cooler ejecta component has not reached CIE yet.

To examine the cooling across the SNR, we map $kT_{\rm init}/kT_{\rm e,1}$ in Figure~\ref{fig:ej1params}. Enhanced ratios of $kT_{\rm init}/kT_{\rm e,1}$ may reflect more recent, rapid cooling (a shorter recombination timescale $t_{\rm rec,1}$) or a lower-density medium. We find the northwest has the highest ratios, and the south and central regions have comparatively lower $kT_{\rm init}/kT_{\rm e,1}$.

As stated in Section~\ref{subsec:fitting}, we tied $kT_{\rm init}$ of the cooler ejecta component to that of the hotter component. We note that the best-fit $kT_{\rm init}$ and $\tau_{1}$ are correlated: greater $kT_{\rm init}$ values correspond to greater $\tau_{1}$ values. This relationship arises since a higher temperature requires more time to reach the current ionization state. To correct for this degeneracy, we map an ``effective $kT_{\rm init}/kT_{\rm e,1}$ ratio'' that is computed by weighting $kT_{\rm init}$ with $\tau_{1}$ and dividing by $kT_{\rm e,1}$. In this case, the trend is even more pronounced; effective $kT_{\rm init}/kT_{\rm e,1}$ is greatest in the northwest and decreases in the east and south directions.

In Figure~\ref{fig:ej1params}, we also present a map of the recombination age $t_{\rm rec,1}$. To compute $t_{\rm rec,1}$, we first estimate the electron density $n_{\rm e}$ in each region across the SNR using the \texttt{norm} parameter of the cooler ejecta component, since \texttt{norm} = (10$^{-14}/4 \pi D^2) \int n_{\rm H} n_{\rm e} dV$, where $D$ is the distance to the SNR, $n_{\rm H}$ is the hydrogen number density and $V$ is the volume of each region. We adopt a distance of $D = 9.3$~kpc, an intermediate value of the distance estimates of 8--11.3~kpc in the literature (e.g., \citealt{brogan01,zhu14,rana18}). We assume that $n_{\rm{H}} \approx 1.2n_{\rm{e}}$, as is the case for a fully ionized plasma with solar abundances. To calculate the volume of each region, we approximate each 0.5\arcmin$\times$0.5\arcmin\ region as a rectangular prism. We assume the SNR is spherical and estimate the depth $l$ of each region, where $l=2\sqrt{{\rm R}_s^2 - r^2}$, $R_{\rm s}$ is the radius of the SNR, and $r$ is the projected distance from the region to the SNR center. We adopt a radius $R_{\rm s} = 2.37$\arcmin, corresponding to 6.6~pc at $D = $9.3~kpc, and a SNR center of right ascension $\alpha =$ 19$^{\rm h}$11$^{\rm m}$7.11$^{\rm s}$ and declination $\delta = +9^{\circ}$6\arcmin14.04\arcsec. Furthermore, we assume electron pressure equilibrium (i.e., $n_{\rm e,ISM}kT_{\rm ISM}$=$n_{\rm e,1}kT_{\rm e,1}$=$n_{\rm e,2}kT_{\rm e,2}$) in order to solve for the unknown filling factors of each plasma component. Finally, to calculate $t_{\rm rec,1}$, we divide the best-fit $\tau_{1}$ by the derived $n_{\rm e,1}$. We note that since the environments of mixed-morphology SNRs like W49B tend to be quite clumpy \citep[e.g., see][and references there within]{zhang19}, this calculation (which assumes a uniform $n_{\rm e}$) likely underestimates the true $n_{\rm e}$ of the environment, leading to an overestimation in $t_{\rm rec}$. Moreover, $t_{\rm rec}$ depends on $f$, the volume filling factor, since $t_{\rm rec} \propto f^{0.5}$. For this calculation, we have assumed that $f_1+f_2+f_3 = 1$. 

In the central and western regions, we find that $t_{\rm rec,1}$ = 600--2000 years, with a gradient of increasing ages toward the east. Assuming an ionization timescale of $\tau_{1} \sim3 \times 10^{12}$ cm$^{-3}$~s (the boundary around which a CIE model fit better than an RP model), the eastern regions in CIE have $t_{\rm rec,1}$ values of $\sim3000-6000$ years. Our results are similar to those of \cite{zhou18}, who found t$_{\rm rec} \approx 1000-4000$ years in the center and west regions and $t_{\rm rec} \gtrsim$ 8,000 years in the east (we note that they used a single RP in their model, reflecting the hotter ejecta). Similarly, \cite{sun20}, whose model of the integrated spectrum included the same components as ours, derived a $t_{\rm rec}$ of 3400$\pm$200 years. 

\begin{figure}
\begin{center}
\includegraphics[width=\columnwidth]{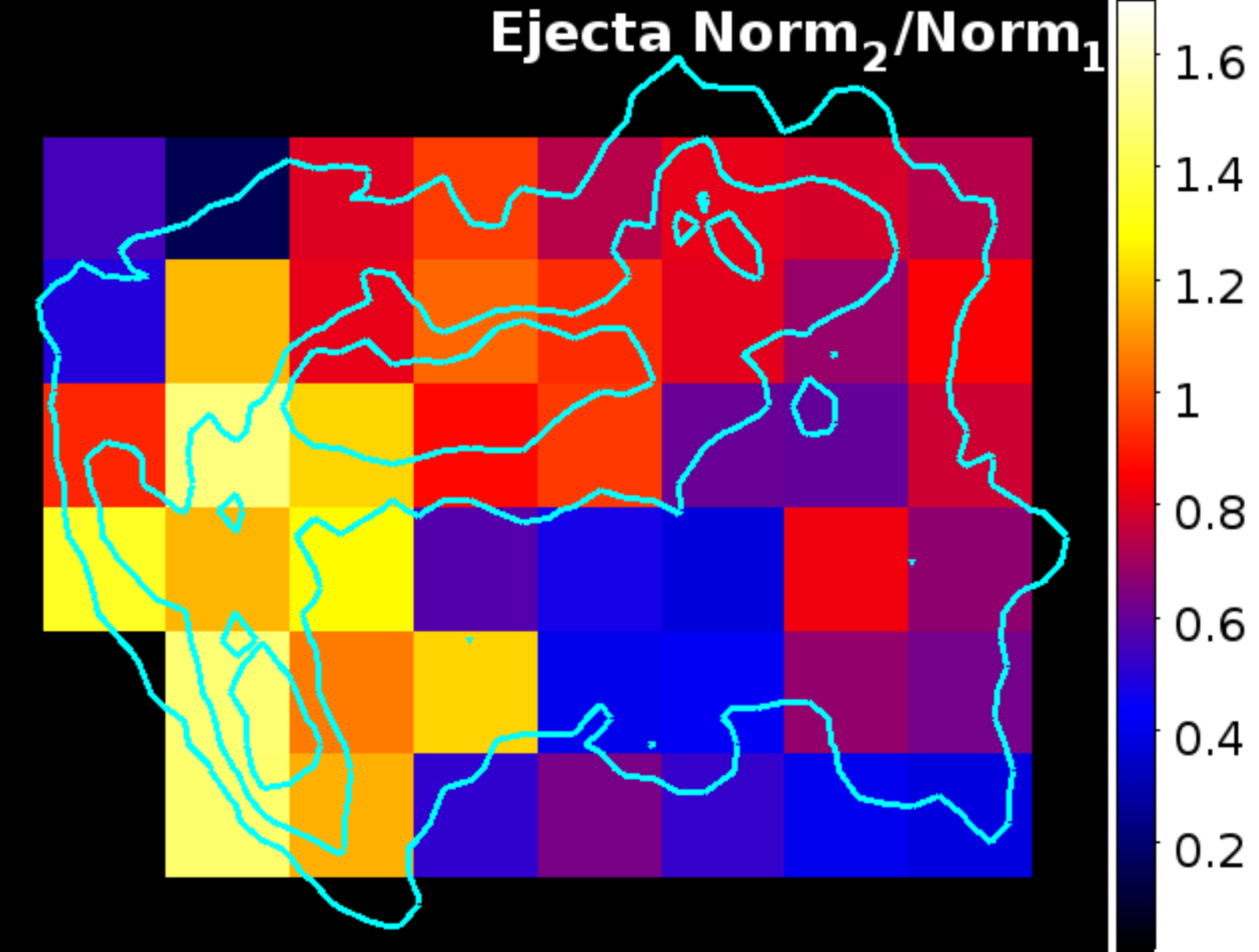}
\end{center}
\caption{Map of the best-fit \texttt{norm$_{2}$}/\texttt{norm$_{1}$}, the ratio of the normalizations of the hotter and cooler ejecta components, with cyan contours overplotted from the broad-band X-ray emission (from {\it Chandra}; \citealt{lopez13b}). Larger ratios correspond to greater flux from the hotter RP component. North is up, and east is left.} 
\label{fig:normratio}
\end{figure}

\subsection{Hotter Ejecta Component Maps}
\label{subsec:ej2_maps}

Figure~\ref{fig:ej2params} shows the best-fit parameters associated with the hotter \texttt{vvrnei} ejecta component. The current electron temperature $kT_{e,2}$ ranges from 0.87--1.62~keV, values similar to the temperatures measured by \citealt{lopez13b} ($\approx$1.1--1.8 keV), \citealt{zhou18} ($\approx$0.7--2.2~keV), and \citealt{sun20} (1.60\err{0.02}{0.01}~keV). The best-fit $kT_{\rm init}$ spans from 2.72--5.07~keV, comparable to the global $kT_{\rm init} =$ 4.54\err{0.17}{0.07}~keV found by \cite{sun20} and the range of $kT_{\rm init} \sim$ 2--5~keV reported by \cite{zhou18}.

The $kT_{e,2}$ and $kT_{\rm init}$ maps show significant gradients going from the southwest to the northeast of W49B. Regions coincident with W49B's ``bar'' (associated with a possible jet; \citealt{lopez13a,gonzalez14}) and eastern regions where the SNR is interacting with a molecular cloud \citep{keohane07,zhu14} have elevated $kT_{\rm e,2}$ and $kT_{\rm init}$. A similar trend is evident in $\tau_{2}$, with longer ionization timescales (of $\tau_{2} \gtrsim 4\times10^{11}$~cm$^{-3}$~s) in the east relative to the west (where $\tau_{2} \lesssim 4\times10^{11}$~cm$^{-3}$~s).

\begin{deluxetable}{lccc}[!t]
\tablecolumns{4}
\tablewidth{0pt} 
\tablecaption{Ionization Temperature Results \label{table:iontemps}} 
\tablehead{ \colhead{Element} & \colhead{Cooler CIE } & \colhead{Cooler RP} &\colhead{Hotter RP}  \\
\colhead{} &\colhead{Ejecta $kT_{\rm z}$ (keV)} &\colhead{ $kT_{\rm z}$ (keV)} &\colhead{$kT_{\rm z}$ (keV)}  }
\startdata
Si & 0.5--0.8 & 0.7--0.9 &  1.1--1.3 \\
S & 0.8--1.0 & 1.0--1.4 & 1.6--1.8  \\
Ar & 1.4--1.8 & 1.9--2.4 & 2.2--2.4  \\
Ca & 2.2--2.4 & 2.4--2.7 & 2.7--3.0  \\
\enddata
\end{deluxetable}

\begin{figure*}
\begin{center}
\includegraphics[width=0.9\textwidth]{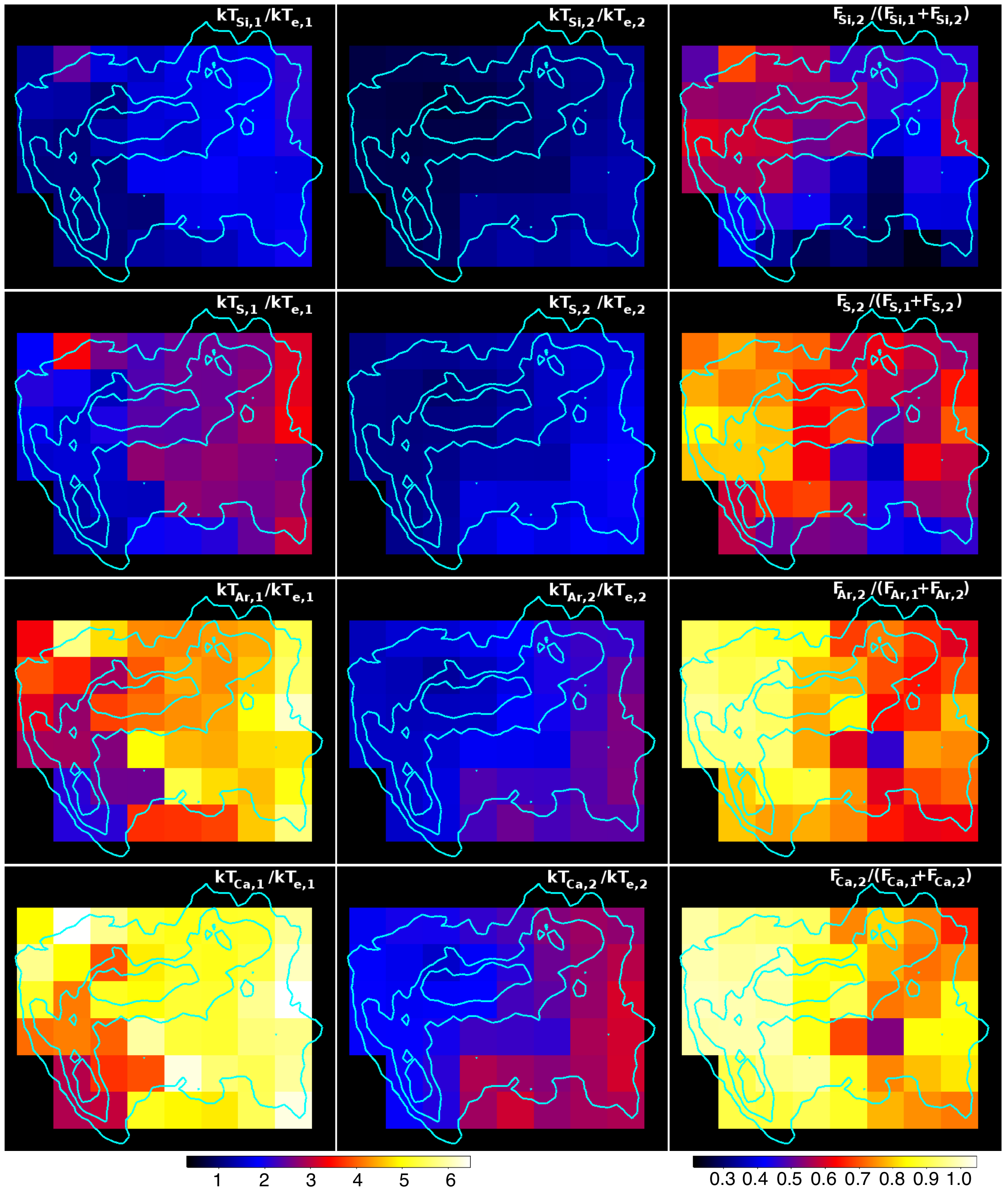}
\end{center}
\vspace{-5mm}
\caption{Maps of the ionization temperature to electron temperature ratios ($kT_{\rm z}/kT_{\rm e}$) for Si, S, Ar, and Ca for both the cooler (left column) and hotter ejecta components (middle column). The $kT_{\rm z}/kT_{\rm e}$ colormaps have the same scale for easier visual comparison. The right column shows the fractional flux of the He-$\alpha$ line of each element in the hot ejecta component (e.g., $F_{\rm Si,1}$) relative to the total from both ejecta components (e.g., $F_{\rm Si,1}+F_{\rm Si,2}$). North is up, and east is left.}
\label{fig:iontemps}
\end{figure*}

As with the cooler ejecta component (see Section~\ref{subsec:ej1_maps}), we assess the efficiency of the rapid cooling by mapping the ratio $kT_{\rm init}/kT_{\rm e,2}$, the $\tau_{2}$-weighted $kT_{\rm init}/kT_{\rm e,2}$, and $t_{\rm rec,2}$ in Figure~\ref{fig:ej2params}. The western regions of W49B exhibit a much larger temperature difference than the eastern regions, consistent with the results for the cooler ejecta component. However, the hotter component has the largest ratios in the southwest, whereas the cooler ejecta had the most cooling in the northwest. 

Our recombination age $t_{\rm rec,2}$ map for the hotter component has a similar east-west gradient as that of $t_{\rm rec,1}$: the central and western regions have $t_{\rm rec,2} \approx 500-1000$ years, whereas the eastern side has t$_{\rm rec,2}\approx 1500-2500$ years. Though these values are consistent with those of \cite{zhou18}, our $t_{\rm rec,2}$ range is less than the recombination age of 6000$\pm$400~years derived by \cite{sun20} for their hotter RP. These disparate results likely arise from differences in the estimates of $n_{\rm e,2}$: \cite{sun20} found a hotter RP density of $\sim$2.1~cm$^{-3}$, whereas our $n_{\rm e,2}$ range from $\sim5-20$~cm$^{-3}$.

Figure~\ref{fig:normratio} shows the ratio of the hotter-to-cooler ejecta best-fit \texttt{norm} values, where \texttt{norm$_{2}$}/\texttt{norm$_{1}$}$>$1 indicates that a region has greater emission measure in the hotter component. This map demonstrates that the hotter ejecta dominates eastern regions of W49B, the cooler ejecta dominates south-western regions, and the two are equally present in the north/northwest regions.

\subsection{Ionization Temperatures}\label{subsec:iontemps}

In Table~\ref{table:iontemps}, we summarize the range of the ionization temperatures $kT_{\rm z}$ obtained from our spectral modeling. Here we divide the findings into three categories: the cooler ejecta in CIE, the cooler overionized ejecta, and the hotter overionized ejecta. For all regions, we find that higher-mass elements exhibit greater $kT_{\rm z}$ than lower-mass elements, suggesting that the former are further out of CIE than the latter elements. This finding is consistent with the fact that lighter elements require shorter ionization timescales to reach CIE at temperatures $\gtrsim$1~keV \citep{smith10}. To determine the degree of overionization and to probe the physical origin of the recombining plasma, we compare $kT_{\rm z}$ of each element to $kT_{\rm e}$ in the ejecta components.

In Figure~\ref{fig:iontemps} we plot the ratio of the ionization temperature to the current electron temperature ($kT_{\rm z,1}/kT_{\rm e,1}$ and $kT_{\rm z,2}/kT_{\rm e,2}$) for the Si, S, Ar, and Ca in both ejecta components. For the cooler plasma in CIE, in the east of W49B, only Si has $kT_{\rm Si,1}/kT_{\rm e,1} \sim$ 1. By comparison, $kT_{\rm S,1}/kT_{\rm e,1} \sim$ 1.1--1.8 in the same regions, and $kT_{\rm Ar,1}$ and $kT_{\rm Ca,1}$ are significantly higher, with $kT_{\rm Ar,1}/kT_{\rm e,1} \sim 3$ and $kT_{\rm Ca,1}/kT_{\rm e,1}\sim4$. Consequently, despite the best-fit model of the eastern cooler ejecta including a CIE component, the elements heavier than Si are still overionized. This result indicates that the plasma has cooled rapidly, and only the Si (and probably elements lighter than Si) has reached CIE. In the central and western regions, all of the elements in the cooler plasma have $kT_{\rm z,1}/kT_{\rm e,1} > 1$ (indicative of overionization) and show gradients of increasing $kT_{\rm z,1}/kT_{\rm e,1}$ from southeast to northwest.

We note that the vast majority of the emission from S, Ar, and Ca is from the hotter ejecta component, whereas the Si emission comes from both ejecta components. Figure~\ref{fig:iontemps} (right column) demonstrates this point by mapping the ratio of the flux in the He-$\alpha$ line of the hotter ejecta $F_{\rm z,2}$ to the total He-$\alpha$ line flux of both ejecta components for each element ($F_{\rm z,1}+F_{\rm z,2}$). Here only a fraction ($\lesssim 60\%$) of the total Si emission is from hotter ejecta component. Given that the Ar and Ca contribute negligibly to the flux in the cooler RP in particular, the $kT_{\rm Ar,1}/kT_{\rm e,1}$ and $kT_{\rm Ca,1}/kT_{\rm e,1}$ maps have large uncertainties and should be interpreted with caution.

For the hotter RP ejecta component, all of the elements show a gradient of increasing $kT_{\rm z,2}/kT_{\rm e,2}$ toward the west/southwest,  consistent with the $kT_{\rm init}/kT_{\rm e,2}$ map in Figure~\ref{fig:ej2params}. In the western and central regions, all elements have $kT_{\rm z,2}/kT_{\rm e,2} > 1$. However, in the eastern regions, although the ratios of $kT_{\rm z,2}/kT_{\rm e,2} \geq$ 1 for S, Ar, and Ca, we find that Si has kT$_{\rm z}$/kT$_{e,2} < 1$, indicating that Si is underionized in the east. We note that the $\sim$60\% of Si emission in the eastern regions comes from the hotter ejecta component, indicating that this measurement is robust.

\subsection{Fe RRC Map} \label{subsec:FeRRC}

\begin{figure}
\begin{center}
\includegraphics[width=\columnwidth]{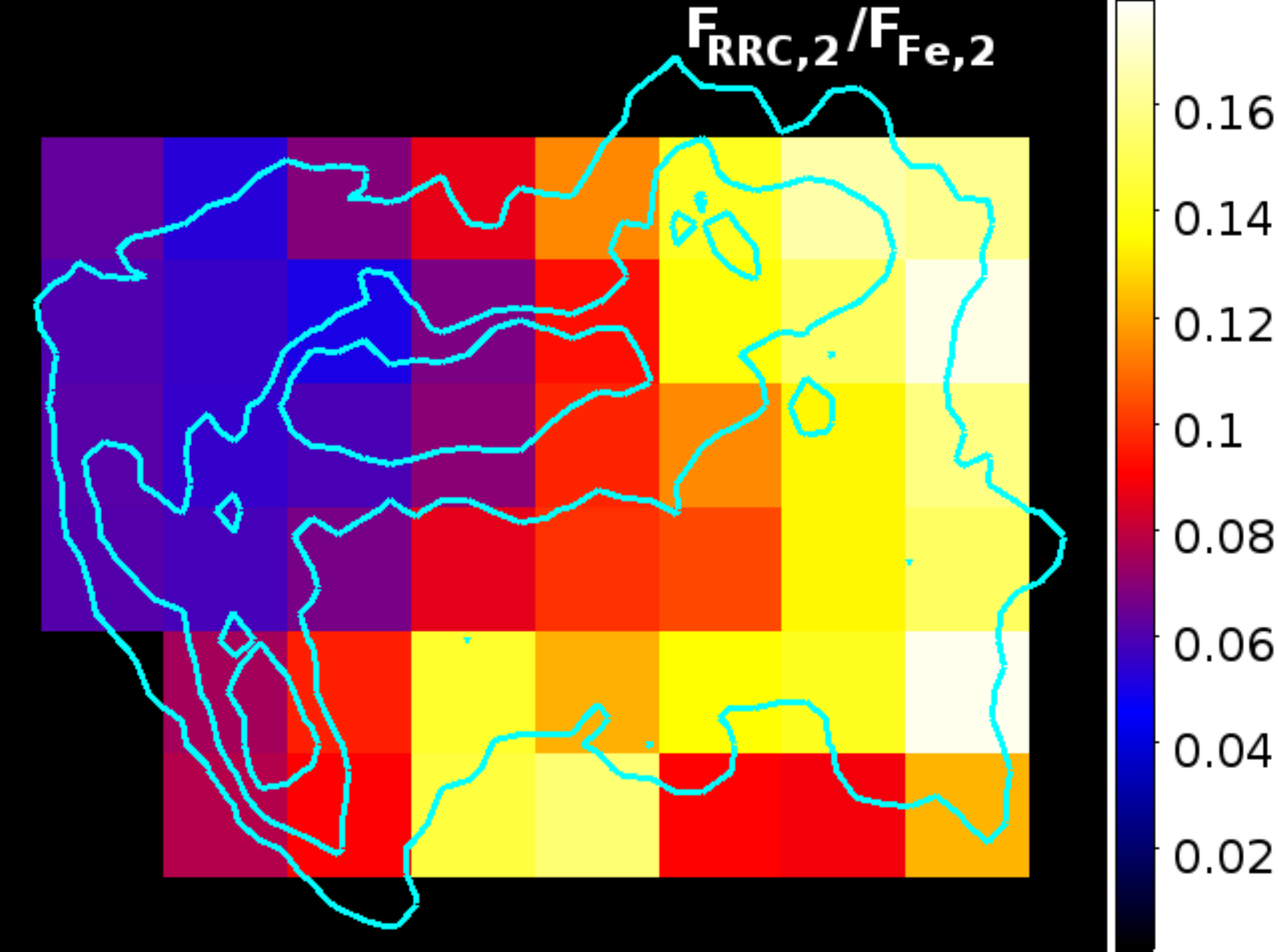}
\end{center}
\caption{Map of the ratio between the continuum-subtracted 8.8--10.0~keV Fe RRC flux ($F_{\rm RRC,2}$) and the 6.3--6.9~keV Fe He-$\alpha$ flux ($F_{\rm Fe,2}$) of the hotter ejecta component. a method of measuring the degree of overionization. North is up, and East is left. } 
\vspace{3mm}
\label{fig:FeRRC}
\end{figure}

The He-like Fe radiative recombination continuum (RRC) line at $E_{\rm edge} =$ 8.83 keV in W49B is a useful spectral feature to constrain further the ionization state of the RP \citep{ozawa09}. Toward this end, we map the ratio of the Fe RRC flux (from 8.8--10~keV; $F_{\rm RRC,2}$) to the Fe He-$\alpha$ flux (from 6.3--6.9~keV; $F_{\rm Fe,2}$) from the hotter ejecta component in Figure~\ref{fig:FeRRC} (the hotter component is responsible for nearly all of the Fe emission). The map matches the results of \cite{yamaguchi18}, who used exposure-corrected {\it NuSTAR images} to show the ratio of the 8.8--10~keV band to the 6.4--6.8~keV band. The spatial distribution of $F_{\rm RRC,2}/F_{\rm Fe,2}$ is enhanced towards the west indicating a higher degree of overionization.

\section{Discussion} \label{sec:discussion}

In this section, we synthesize the results presented in Section~\ref{sec:results} to draw conclusions about the presence and origin of the recombining plasma in W49B. We find that W49B exhibits signatures of overionization throughout the entire SNR, in contrast to previous spatially-resolved studies (e.g., \citealt{miceli10, lopez13b, zhou18}) that identified RP in the west and central regions and no evidence of RP in the east. The exceptions are \cite{yamaguchi18} who found low degrees of overionization in the east using {\it NuSTAR} data and \cite{sun20} who analyzed the global {\it XMM-Newton} spectra and found that the best-fit model had all of the ejecta in an overionized state.

We find a gradient of increasing overionization from east to west for both the cooler (with $kT_{e,1}=0.31-0.76$~keV) and hotter (with $kT_{\rm e,2}=0.87-1.62$~keV) ejecta components, based on the maps of $kT_{\rm init}/kT_{\rm e}$, $\tau$, $t_{\rm rec}$ (Figures~\ref{fig:ej1params} and ~\ref{fig:ej2params}), $kT_{\rm z}/kT_{\rm e}$ of each element (Figure~\ref{fig:iontemps}), and the Fe RRC to He-$\alpha$ flux ratio (Figure~\ref{fig:FeRRC}). Given that W49B is impeded by molecular material to its east/southeast \citep{keohane07,lopez13a,zhu14}, whereas it is now expanding into a lower-density ISM in the west, our results are consistent with the overionization predominantly arising from adiabatic expansion, as suggested in previous work \citep{miceli10, lopez13b, yamaguchi18, zhou18}.

\subsection{RP in Western and Central Regions}

Although both ejecta components are generally more overionized in the west, the cooler RP is most strongly overionized in the northwest (Figure~\ref{fig:ej1params}), and the bulk of the hotter RP overionization is found in the southwest (Figure~\ref{fig:ej2params}). Both results are consistent with the adiabatic expansion scenario, as described above. However, it is possible that the southwest (instead of west/northwest) enhancement of overionization in the hot RP arises from interaction with CSM material there \citep{keohane07}, though that material may be in foreground \citep{lacey01}.

We again note that a majority of the emission in the southwest of W49B (see Figure~\ref{fig:normratio}) arises from the cooler component, whereas the emission in the northwest is produced equally by the hot and cool RPs. Taken together with the maps of Figures~\ref{fig:ej1params} and~\ref{fig:ej2params}, we see that the bulk of the overionized ejecta is toward the west/northwest, farthest away from the eastern molecular cloud and consistent with rapid cooling from adiabatic expansion.

For these central and western regions, the $kT_{\rm z}/kT_{\rm e}$ maps for Si, S, Ar, and Ca show increasing ionization temperatures for higher-mass elements (see Figure~\ref{fig:iontemps}). This increasing ionization temperature is similar to that found in past studies \citep{kawasaki05,lopez13b,sun20}, consistent with the fact that heavier elements require a factor of a few higher ionization timescales to reach CIE at temperatures of $\gtrsim$1~keV \citep{smith10}. These ionization temperatures are higher in the western regions of W49B, supporting the existence of more highly ionized plasma there. As each element's ionization temperature map shows a similar distribution, we conclude that the different ionization temperatures between Si/S and Ar/Ca do not arise because of associations with distinct RP components. 

\subsection{RP in Eastern Regions}

We find that the cooler ejecta in the east of W49B, where the SNR is impeded by molecular material, is best fit with a CIE component. This finding is consistent with the results of \cite{zhou18}, who found that their single ejecta component is in CIE to the east. This result indicates that either the ejecta cooled more rapidly or for longer times than the other parts of the SNR, or that this part of the SNR was never overionized. As we are unable to constrain $t_{\rm rec,1}$ in these regions, we use the $kT_{z,1}/kT_{e,1}$ ratios in Figure \ref{fig:iontemps} to constrain its ionization history. We find that nearly all elements have $kT_{z,1}/kT_{e,1} >1$, suggesting that the plasma was overionized and is progressing toward CIE. The ratio $kT_{z,1}/kT_{e,1}$ is smaller for the lighter elements, with $kT_{z,1}/kT_{e,1} \sim 1$ for Si, consistent with lighter elements requiring shorter $\tau$ to reach CIE \citep{smith10}. However, we note that the majority of the flux in the east of W49B is from the hotter RP (see Figure~\ref{fig:normratio}), suggesting that the majority of ejecta in the east is hotter and is still overionized.

In contrast to \cite{lopez13b} and \cite{zhou18}, we find evidence of overionization in the hotter ejecta in the east of W49B where the SNR is impacting molecular material, though the amount of overionization is $\gtrsim$1.5--2$\times$ less than in the west. These differences suggest that the regions closest to the molecular cloud in the east did not cool as rapidly as the freely expanding regions in the west/northwest and the less dense regions in the southwest. This reduced cooling may result from the lack of low-density medium to expand into or because cooling via thermal conduction is not as efficient as cooling from adiabatic expansion.

\subsection{Thermal Conduction Timescales}

As discussed above, we find greater overionization in the west of W49B, but overionized ejecta is present in the east as well. In these eastern regions, the cooler ejecta is partially in CIE ($kT_{\rm Si,1}/kT_{\rm e,1}\approx1$ whereas other elements have $kT_{\rm z,1}/kT_{\rm e,1}>1$) and the hotter ejecta is entirely overionized. While the western regions' cooling is likely a result of adiabatic expansion, the overionization in the eastern regions could be a result of adiabatic expansion and/or thermal conduction.

To examine the origin of the rapid cooling in the east, we calculate the thermal conduction timescale $t_{\rm cond}$ using

\begin{equation}
t_{\rm cond} \approx 634 \bigg( \frac{n_{\rm e}}{{\rm 1~cm}^{-3}} \bigg) \bigg( \frac{\ell_{\rm T}}{1~{\rm pc}} \bigg)^{2} \bigg( \frac{kT_{\rm e}}{1.0~{\rm keV}} \bigg)^{-5/2} \bigg( \frac{\rm{ln} \Lambda}{32} \bigg)~{\rm yr}
\label{eq:tcond}
\end{equation}

\noindent 
where $n_{\rm e}$ is the average electron density, $\ell_{T}$ is the scale length of the temperature gradient, $kT_{\rm e}$ is the average temperature in these regions, and ln $\Lambda$ is the Coulomb logarithm \citep{spitzer62,kawasaki02,zhou14}. Using Equation~\ref{eq:tcond}, we estimate $t_{\rm cond}$ of the eastern regions for both ejecta components and compare these estimates to t$_{\rm rec}$. If $t_{\rm cond} <t_{\rm rec}$, then it is plausible that thermal conduction is the origin of the RP. Otherwise, an alternate cooling mechanism (e.g., adiabatic expansion) is necessary. 

In the eastern regions, we find $n_{\rm e,2}\approx10$~cm$^{-3}$ for the hotter ejecta and $n_{\rm e,1}\approx30$~cm$^{-3}$ for the cooler ejecta (as calculated in Section~\ref{subsec:ej1_maps}).  We assume $\ell_{\rm T}=3$~pc ($\sim$1.1\arcmin) to match the distance from the CIE boundary to the unshocked, cool ISM east of the SNR.  Using the average current electron temperatures of $kT_{e,1}\approx0.6$~keV and $kT_{e,2}\approx1.6$~keV, we find t$_{\rm cond, 1}= 613$~kyr for the cooler RP and $t_{\rm cond, 2}=17.6$~kyr for the hotter RP. These thermal conduction timescales are both significantly greater than the estimated $t_{\rm rec}\sim 1500-6000$~years, and thus thermal conduction is unlikely the source of cooling. 

As noted in the introduction, models \citep{zhou11,zhang19} find that thermal conduction is necessary to reproduce the observed morphology (specifically the bar) of W49B. These works split thermal conduction into large- and small-scale processes, claiming that large-scale thermal conduction smoothed the temperature and density distributions, whereas the small-scale thermal conduction led to cloud evaporation (see \citealt{cowie81} and \citealt{white91}) that produced a thermal X-ray emitting core and overionization features. We note that past studies \citep{vink12,zhou14,sun20} have set $\ell_{\rm T} \sim R_{\rm SNR}$, the radius of the SNR, thereby calculating the timescale for large-scale thermal conduction.

To investigate the timescale for small-scale thermal conduction, we compute the necessary $\ell_{\rm T}$ in order for $t_{\rm cond} <t_{\rm rec}$. We perform this test on the hotter RP (with $kT_{\rm e,2}\approx1.6$~keV and $n_{\rm e,2}\approx10$~cm$^{-3}$) as it dominates the flux in the eastern regions (see Figure~\ref{fig:normratio}). For this eastern ejecta, we find that $\ell_{\rm T}\lesssim1$~pc in order for $t_{\rm cond} \lesssim t_{\rm rec,2} \approx$ 2000 years. Consequently, dense ISM clouds of sizes $\lesssim$1~pc are necessary in order for thermal conduction to account for the observed cooling. We can further constrain these cloud sizes through the thermal conductivity requirement that the scale length $\ell_{\rm T}$ is greater than the electron mean free path $\lambda_{\rm e} \approx 0.1\ (kT_{\rm e}/0.6  {\rm \ keV})^2\ (n_{\rm e}/10 {\rm \ cm}^{-3})^{-1}$ pc \citep{cowie77}, which gives us a lower limit on the scale length. Thus, we find possible values for the clumpy ISM cloud length scale of $0.1 \lesssim \ell_{\rm T} \lesssim 1$ pc for the hotter ejecta. Otherwise, rapid cooling via thermal conduction is not a plausible origin of the eastern hotter overionized plasma. We note that \cite{mckee77} estimated cloud radii of 0.4--10~pc, and our limits are on the lower end of their predictions.

\section{Conclusions} \label{sec:conclusions}

We performed a spatially-resolved study using deep {\it XMM-Newton} observations of W49B to investigate the presence, location, and physical origin of overionized plasma within the SNR. To that end, we modeled the spectra of 46 0.5\arcmin $\times$ 0.5\arcmin\ regions in W49B. We make use of the high signal to fit the data with a 3-component model: one ISM component plus two ejecta components. To investigate the degree of overionization, we produced temperature, $\tau$, and $t_{\rm rec}$ maps for each ejecta component as well as ionization temperature to current temperature maps for Si, S, Ar, and Ca.

We find that W49B contains overionized plasma across the entire SNR, present in a gradient of increasing overionization from east to west. Our results are broadly consistent with past studies of recombining plasma in W49B (e.g., \citealt{miceli10,lopez13b,yamaguchi18,zhou18,sun20}). Given that the western regions furthest from the eastern molecular cloud interaction \citep{keohane07,zhu14} show the greatest overionization (with $kT_{\rm init}/ kT_{\rm e} \approx$ 4),  we attribute the origin of the majority of recombining plasma in W49B to rapid cooling from adiabatic expansion of shock-heated plasma into a lower-density ISM. In contrast with the results of \cite{lopez13b} and \cite{zhou18}, we find significant overionization ($kT_{\rm init}/kT_{\rm e} \approx$ 2.7) in the eastern regions of the SNR as well, mainly from the hotter ejecta.

Given that the eastern regions are interacting with a molecular cloud, rapid cooling via thermal conduction is a possible origin for this eastern plasma (as suggested by \citealt{sun20} for their hotter ejecta component). For our hotter ejecta, we find that in order for the thermal conduction timescale to be less than the recombination timescale, t$_{\rm cond} <$ t$_{\rm rec}\approx2000$ yrs, the temperature gradient length scales must be $\ell_{\rm T}\lesssim$ 1~pc. The requirement that the length scale be greater than the electron mean free path sets a lower limit of 0.1~pc. These values place constraints on the clumpy ISM/CSM cloud size required for small-scale thermal conduction via cloud evaporation (as opposed to large-scale thermal conduction; \citealt{zhou11,zhang19}) to be a plausible origin of the overionized plasma in the eastern regions. Given theoretical cloud size estimates of 0.4--10~pc \citep{mckee77}, we find that small-scale thermal conduction via cloud evaporation can explain the existence of the hotter, eastern overionized plamsa.

Our spatially-resolved spectral analysis of W49B, combined with the previous studies of W49B's abundances, recombining plasma, and morphology (e.g., \citealt{kawasaki05,lopez09,ozawa09,yang09,miceli10,lopez13b,lopez13a,zhu14,yamaguchi18,sun20}) can be used to inform future simulations investigating the progenitor, explosion processes, and recombination physics required to create this SNR. For example, simulations by \citep{zhou11,zhang19} investigate the recombination physics and ISM structure required to produce W49B's morphology assuming a spherically-symmetric explosion and concluded that small-scale thermal conduction via cloud evaporation is necessary to reproduce its features. Our observational study confirms that small-scale thermal conduction is a viable origin for RP in the east of W49B regardless of whether the explosion was symmetric or asymmetric.

\acknowledgments

LAL is supported by a Cottrell Scholar Award from the Research Corporation of Science Advancement. Parts of this research were supported by the Australian Research Council Centre of Excellence for All Sky Astrophysics in 3 Dimensions (ASTRO 3D), through project number CE170100013. KAA is supported by the Danish National Research Foundation (DNRF132). 

\software{{\it XMM-Newton} SAS (v15.0.0; \citealt{gabriel04}), XSPEC (v12.9.0; \citealt{arnaud96}), ATOMDB v3.0.9; \citealt{smith01,foster12}, ftools \citep{blackburn95}}

\begin{appendix}


\begin{deluxetable*}{lllllllllll} \rotate
\tablenum{2}
\tabletypesize{\footnotesize}
\tablecolumns{11}
\tablewidth{0pt} 
\tablecaption{Best-fit Model Parameters\label{table:fits}} 
\tablehead{ \colhead{Parameter} & \colhead{Region 1}& \colhead{Region 2}& \colhead{Region 3}& \colhead{Region 4}& \colhead{Region 5}& \colhead{Region 6}& \colhead{Region 7}& \colhead{Region 8}& \colhead{Region 9}& \colhead{Region 10}}
\startdata 
$N_{\rm{H}}$ (10$^{22}$ cm$^{-2}$)      &7.57\err{0.04}{0.04} 	&7.51\err{0.06}{0.04} 	&7.595\err{0.04}{0.02} 	&7.46\err{0.04}{0.09} 	&7.62\err{0.03}{0.03} &7.86\err{0.04}{0.04} &7.67\err{0.06}{0.06}	 &7.32\err{0.22}{0.08}  &8.11\err{0.05}{0.05} 	 &8.10\err{0.05}{0.02}     \\
\cutinhead{ISM \texttt{vapec}\tablenotemark{a} component} 
$kT_{\rm e}$ (keV) 	             	&0.154\err{0.003}{0.004}&0.173\err{0.003}{0.004}&0.187\err{0.002}{0.001}&0.211\err{0.007}{0.020}&0.1983\err{0.001}{0.001}&0.1746\err{0.001}{0.001}	&0.178\err{0.001}{0.001} &0.188\err{0.006}{0.006} &0.177\err{0.001}{0.001}&0.182\err{0.001}{0.001}  \\
Redshift\tablenotemark{b} ($10^{-3}$)                &3.12 	&$-$0.72 	&-3.14 	&$-$3.33 	&3.38	&$-$5.43 	&-3.65 	 &4.03  &$-$0.24  &$-$2.99        \\
Norm\tablenotemark{b} 	(cm$^{-5}$)             &1.81			&0.77 			&1.94 			&1.13	&2.81 			&8.45 	&4.76			 &1.75  &2.33 			 &4.66     \\
\cutinhead{Ejecta 1 \texttt{vvapec} or \texttt{vvrnei}\tablenotemark{c} component\tablenotemark{d}}
$kT_{\rm e}$ (keV)        			&0.45\err{0.03}{0.09} 			&0.31\err{0.02}{0.02}&0.47\err{0.03}{0.02}&0.52\err{0.03}{0.04}&0.53\err{0.02}{0.01}&0.51\err{0.01}{0.01}&0.51\err{0.01}{0.02}	 &0.44\err{0.02}{0.01}&0.40\err{0.03}{0.02}&0.53\err{0.02}{0.03}     \\
$\tau_{1}$ ($10^{11}$ cm$^{-3}$ s) & --- 			& --- 			&10.3\err{0.6}{0.6}  	&8.86\err{1.10}{0.58} 	&6.49\err{0.21}{0.21} 	&8.35\err{0.23}{0.17} 	&6.57\err{0.27}{0.29}	 &5.42\err{0.43}{0.40}  & --- 			 &14.2\err{1.3}{0.7}        \\
Redshift\tablenotemark{b} ($10^{-3}$)          &$-$2.48 	&$-$2.35	&$-$0.09 	&0.05 	&$-$4.27 	&0.25 &$-$0.07	 &$-$3.35 &$-$4.17  &$-$2.68    \\
Norm ($10^{-3}$ cm$^{-5}$)    &3.91\err{0.34}{0.30} 	&14.7\err{6.3}{0.8} 	&4.36\err{0.34}{0.71} 	&3.67\err{0.53}{0.45} 	&6.65\err{0.38}{0.29} 	&8.71\err{0.17}{0.19} 	&6.12\err{0.30}{0.31} &3.01\err{0.37}{0.43}  &5.08\err{1.65}{0.66} 	 &4.69\err{0.42}{0.37}       \\
\cutinhead{Ejecta 2 \texttt{vvrnei}\tablenotemark{e} component}
$kT_{\rm e}$ (keV)                  &1.64\err{0.02}{0.02} &1.47\err{0.02}{0.02} &1.48\err{0.03}{0.01}&1.35\err{0.02}{0.02} &1.28\err{0.03}{0.01} &1.13\err{0.07}{0.01}&1.07\err{0.08}{0.08} &1.07\err{0.02}{0.03}&1.59\err{0.02}{0.02} &1.68\err{0.01}{0.01}     \\
$kT_{\rm init}$ (keV)         &5.07\err{0.41}{0.19} 	&4.05\err{0.14}{0.14} 	&3.87\err{0.06}{0.08} 	&3.75\err{0.08}{0.16} 	&4.10\err{0.18}{0.16} 	&3.86\err{0.10}{0.03} &4.06\err{0.21}{0.08}	 &4.06\err{0.42}{0.14}  &4.43\err{0.28}{0.21} 	 &3.86\err{0.05}{0.10}   \\
Mg	                        &$<$1.85			&$<$2.33		 	&5.18\err{0.76}{0.57} 	&6.51\err{0.99}{1.16} 	&4.80\err{0.43}{0.46} 	&4.48\err{0.66}{0.49} 	&5.15\err{0.87}{0.28}	 &3.57\err{0.85}{0.72}  &8.43\err{0.73}{0.70} 	 &6.89\err{0.56}{0.81}        \\
Si		                  &6.67\err{0.19}{0.26} 	&11.7\err{0.3}{0.3} 	&9.18\err{0.29}{0.18} 	&9.89\err{0.66}{0.52} 	&9.66\err{0.40}{0.40} 	&11.7\err{0.4}{0.2} 	&11.8\err{0.4}{0.5}	 &12.0\err{1.1}{1.0}  &13.2\err{0.3}{0.3}  &9.8\err{0.2}{0.2}          \\
S 	                        &8.36\err{0.19}{0.18} 	&14.7\err{0.2}{0.3} 	&12.4\err{0.3}{0.2} 	&14.1\err{0.7}{0.6} 	&13.3\err{0.4}{0.5} 	&15.1\err{0.2}{0.2} 	&16.9\err{0.7}{0.8}	 &17.1\err{1.2}{1.1}  &17.4\err{0.2}{0.2} 	 &13.0\err{0.2}{0.3}          \\
Ar	                        &11.4\err{0.5}{0.5} 	&19.3\err{0.8}{0.8} 	&14.7\err{0.5}{0.5} 	&16.5\err{0.5}{0.5} 	&15.6\err{0.4}{0.5} 	&17.8\err{0.5}{0.5} 	&20.4\err{0.5}{0.6}	 &18.8\err{0.7}{0.7}  &19.5\err{0.6}{0.6} 	 &14.7\err{0.4}{0.5}         \\
Ca	                        &11.4\err{0.4}{0.5} 	&18.9\err{0.6}{0.6} 	&14.8\err{0.5}{0.4} 	&15.5\err{0.5}{0.5} 	&14.1\err{0.3}{0.4} 	&17.1\err{0.4}{0.5} 	&20.8\err{0.9}{0.5}	 &17.9\err{0.7}{1.4}  &19.1\err{0.6}{0.6} 	 &14.6\err{0.3}{0.3}         \\
Cr	                        &19.2\err{2.8}{3.0} 	&31.1\err{4.4}{4.3} 	&18.6\err{2.9}{2.7} 	&28.6\err{3.4}{3.3} 	&27.9\err{2.9}{2.8} 	&25.2\err{2.8}{1.9} 	&22.2\err{3.5}{3.2}	 &30.1\err{6.4}{5.6} 	 &32.5\err{3.4}{3.8} 	 &15.1\err{1.9}{1.9}         \\
Mn	                        &59.7\err{5.1}{9.2} 	&77.8\err{9.9}{11.4} 	&32.7\err{4.7}{8.9} 	&55.7\err{7.8}{9.0} 	&48.8\err{5.4}{6.8} 	&69.8\err{12.9}{11.8} 	&38.9\err{9.4}{9.6}	 &39.8\err{20.5}{14.7}  &82.6\err{9.03}{10.4} 	 &32.8\err{3.8}{4.3}          \\
Fe	                        &16.4\err{0.2}{0.5} 	&35.8\err{0.9}{0.7} 	&20.8\err{0.6}{1.1} 	&19.0\err{1.1}{0.7} 	&14.9\err{0.6}{0.4} 	&14.2\err{0.2}{0.7} 	&13.8\err{0.9}{0.7}	 &12.7\err{1.4}{1.1}  &29.0\err{0.9}{0.2}   &20.7\err{0.5}{0.3}         \\
Ni	                        &29.0\err{3.9}{3.8} 	&86.0\err{16.8}{12.9} 	&42.8\err{5.5}{3.6} 	&44.2\err{9.6}{4.7} 	&15.5\err{3.4}{3.4} 	&14.1\err{4.9}{4.4} 	&10.8\err{5.0}{5.1}	 &$<$5.16		 	 &68.1\err{4.6}{4.8} 	 &29.4\err{2.5}{2.3}          \\
$\tau_{2}$ ($10^{11}$ cm$^{-3}$ s) &5.98\err{0.25}{0.23} 	&4.89\err{0.27}{0.29} 	&3.89\err{0.36}{0.08} 	&3.396\err{0.29}{0.33} 	&3.59\err{0.23}{0.09} 	&3.06\err{0.12}{0.08} 	&3.0\err{0.21}{0.08}	 &3.13\err{0.47}{0.21}  &5.55\err{0.30}{0.544} &4.17\err{0.27}{0.28}       \\
Redshift\tablenotemark{b} ($10^{-3}$)          & $-$3.30 	&$-$3.3 	&$-$3.31 	&$-$3.31 	&$-$2.82 	&$-$2.90 	&$-$3.17	 &$-$2.81 &$-$3.22  &$-$3.30     \\
Norm\tablenotemark{f} ($10^{-3}$ cm$^{-5}$)  	&2.14			&2.14 			&3.49 			&3.50 			&4.89 			&7.10 			&4.82			 &2.21 			 &2.49			 &5.46                 \\
$\chi^2$/dof	                  &1.09 			&1.12 			&1.08	 		&1.06 			&1.11 			&1.21 			& 1.14		 &1.15 			 &1.12 			 &1.17                     \\
\enddata
\end{deluxetable*}

\begin{deluxetable*}{lllllllllll} \rotate
\tablenum{2}
\tabletypesize{\footnotesize}
\tablecolumns{11}
\tablewidth{0pt} 
\tablecaption{Best-fit Model Parameters\label{table:fits}} 
\tablehead{ \colhead{Parameter} & \colhead{Region 11}& \colhead{Region 12}& \colhead{Region 13}& \colhead{Region 14}& \colhead{Region 15}& \colhead{Region 16}& \colhead{Region 17}& \colhead{Region 18}& \colhead{Region 19}& \colhead{Region 20}}
\startdata 
$N_{\rm H}$ (10$^{22}$ cm$^{-2}$)      &8.76\err{0.03}{0.07}	 & 7.92\err{0.02}{0.01}	 &8.11\err{0.03}{0.04}&8.01\err{0.01}{0.01}  &7.65\err{0.05}{0.04}	 &7.27\err{0.03}{0.05} &7.90\err{0.06}{0.04}	 &8.00\err{0.02}{0.01} &8.16\err{0.02}{0.01}	 &8.07\err{0.02}{0.01}     \\
\cutinhead{ISM \texttt{vapec}\tablenotemark{a} component} 
$kT_{\rm e}$ (keV) 	                  &0.171\err{0.001}{0.001} &0.176\err{0.001}{0.001}&0.173\err{0.001}{0.001}&0.178\err{0.001}{0.001}  &0.181\err{0.003}{0.001}&0.195\err{0.002}{0.003}&0.175\err{0.001}{0.001}&0.179\err{0.001}{0.001}&0.174\err{0.001}{0.001}	 &0.176\err{0.001}{0.001}  \\
Redshift\tablenotemark{b} ($10^{-3}$)                &3.56	 &3.42	 &2.92	 &3.95	 &$-$0.56	 &4.02	 &5.34	 &$-$3.24	 &$-$5.23	 &$-$0.28        \\
Norm\tablenotemark{b} 	(cm$^{-5}$)             &26.2	 &10.4			 &15.41	 &9.86		 &4.19			 &1.37		 &4.71		 &6.88			 &12.31		 &12.66      \\
\cutinhead{Ejecta 1 \texttt{vvapec} or \texttt{vvrnei}\tablenotemark{c} component\tablenotemark{d}}
$kT_{\rm e}$ (keV)        			&0.60\err{0.01}{0.02} &0.55\err{0.01}{0.02}&0.51\err{0.01}{0.01}&0.52\err{0.01}{0.01}&0.50\err{0.01}{0.01}&0.43\err{0.01}{0.01}	 &0.45\err{0.03}{0.03}&0.54\err{0.01}{0.03}&0.51\err{0.01}{0.02} &0.53\err{0.02}{0.02}     \\
$\tau_{1}$  ($10^{11}$ cm$^{-3}$ s) &7.31\err{0.41}{0.09}	 &7.59\err{0.11}{0.34}	 &7.47\err{0.10}{0.19}	 &6.86\err{0.01}{0.26}&6.52\err{0.16}{0.16}	 &6.08\err{1.49}{0.34}	 &---		 &---		 &14.16\err{0.54}{0.37}	 &7.78\err{0.08}{0.15}        \\
Redshift\tablenotemark{b} ($10^{-3}$)          &$-$4.42	 &$-$4.45	 &$-$4.34	 &$-$4.28  &$-$0.03	 &$-$0.29	 &$-$7.27	 &$-$2.61	 &$-$0.07	 &$-$4.36    \\
Norm ($10^{-3}$ cm$^{-5}$)    &8.06\err{0.15}{0.66}	 &8.49\err{0.19}{0.20} &10.8\err{0.1}{0.2}	 &10.2\err{0.5}{1.0}	 &9.44\err{0.36}{0.21}	 &3.59\err{0.05}{0.59}	 &4.23\err{0.76}{0.76}	 &4.30\err{0.58}{0.17}	 &9.05\err{0.14}{0.14}	 &11.1\err{0.1}{0.4}       \\
\cutinhead{Ejecta 2 \texttt{vvrnei}\tablenotemark{e} component}
$kT_{\rm e}$ (keV)                    &1.88\err{0.02}{0.03} &1.62\err{0.02}{0.01}&1.40\err{0.01}{0.02}&1.17\err{0.01}{0.03}&1.08\err{0.01}{0.07}	 &0.94\err{0.01}{0.02} &1.53\err{0.01}{0.02} &1.61\err{0.01}{0.02} &1.59\err{0.01}{0.03} &1.58\err{0.01}{0.01}     \\
$kT_{\rm init}$ (keV)         &4.26\err{0.10}{0.08} &3.96\err{0.03}{0.05}&4.18\err{0.04}{0.03} &3.94\err{0.11}{0.05}	 &3.97\err{0.09}{0.08}   &4.07\err{0.12}{0.24}	 &4.54\err{0.31}{0.16}	 &4.08\err{0.10}{0.16}	 &4.00\err{0.04}{0.022}	 &3.92\err{0.02}{0.06}    \\
Mg	                        &9.57\err{0.72}{0.54}	 &4.06\err{0.44}{0.61}	 &2.99\err{0.37}{0.40}	 &3.65\err{0.44}{0.34}	 &3.64\err{0.28}{0.52}	 &4.93\err{0.52}{0.49}	 &6.22\err{0.95}{0.77}	 &9.07\err{0.72}{0.56}	 &3.77\err{0.31}{0.34}	 &4.22\err{0.62}{0.28}        \\
Si		                  &11.4\err{0.2}{0.2}	 &9.19\err{0.16}{0.14}	 &9.09\err{0.11}{0.11}	 &11.4\err{0.2}{0.1}	 &9.78\err{0.20}{0.36}	 &11.1\err{0.5}{0.3}	 &12.8\err{0.2}{0.3}	 &13.3\err{0.2}{0.2}	 &9.64\err{0.12}{0.15}	 &10.57\err{0.2}{0.1}          \\
S 	                        &14.2\err{0.2}{0.1}	 &12.4\err{0.1}{0.12}	 &11.5\err{0.2}{0.1}	 &15.7\err{0.2}{0.2}	 &14.3\err{0.5}{0.5}	 &16.0\err{0.4}{0.2}	 &15.7\err{0.3}{0.3}	 &16.2\err{0.1}{0.2}	 &12.1\err{0.1}{0.1} &13.8\err{0.2}{0.1}          \\
Ar	                        &14.5\err{0.3}{0.3}	 &13.9\err{0.3}{0.3}	 &12.3\err{0.2}{0.3}	 &17.8\err{0.4}{0.4}	 &17.0\err{0.4}{0.4}	 &17.4\err{0.7}{0.6}	 &16.5\err{0.6}{0.6}	 &17.2\err{0.4}{0.4}	 &13.93\err{0.2}{0.3}	 &15.4\err{0.4}{0.2}         \\
Ca	                        &14.7\err{0.2}{0.3}	 &13.6\err{0.3}{0.3}	 &12.4\err{0.2}{0.3}	 &17.8\err{0.3}{0.4}	 &18.0\err{0.6}{0.4}	 &17.8\err{0.7}{0.6}	 &17.3\err{0.5}{0.5}	 &17.6\err{0.3}{0.3}	 &14.0\err{0.2}{0.2}	 &15.6\err{0.4}{0.2}         \\
Cr	                        &13.8\err{1.4}{1.3}	 &19.4\err{1.5}{1.4}	 &22.2\err{1.4}{1.8}	 &30.0\err{2.5}{2.5}	 &20.7\err{2.7}{2.6}	 &29\err{5.3}{5.2}	 &30.9\err{2.9}{3.0}	 &21.7\err{2.2}{2.0}	 &16.5\err{1.3}{1.4}	 &23.0\err{1.5}{1.7}         \\
Mn	                        &21.3\err{2.7}{3.2}	 &30.5\err{2.9}{3.0}	 &34.3\err{3.}{3.6}	 &39.6\err{6.5}{4.8}	 &26.7\err{7.8}{7.8}	 &98.5\err{35.8}{34.5}	 &50.3\err{6.9}{6.8}	 &38.3\err{4.6}{4.4}	 &26.6\err{3.4}{2.8}	 &30.9\err{4.4}{2.1}          \\
Fe	                        &17.4\err{0.2}{0.1}	 &18.2\err{0.1}{0.4}	 &16.3\err{0.1}{0.1}	 &15.3\err{1.2}{0.5}	 &11.6\err{0.6}{0.6}	 &13.1\err{0.9}{0.6}	 &24.5\err{0.8}{0.5}	 &25.7\err{0.3}{0.1}	 &18.5\err{0.3}{0.3}	 &20.6\err{0.4}{0.1}         \\
Ni	                        &21.0\err{1.9}{1.8}	 &38.8\err{2.1}{2.3}	 &31.4\err{2.8}{1.9}	 &28.8\err{3.0}{3.1}	 &15.1\err{3.2}{4.3}	 &$<$3.67			 &62.6\err{5.3}{6.1}	 &61.4\err{2.1}{4.4}	 &39.7\err{1.8}{1.8}	 &46.0\err{2.9}{1.8}          \\
$\tau_{2}$ ($10^{11}$ cm$^{-3}$ s) &5.01\err{0.09}{0.32}	 &3.96\err{0.21}{0.12}	 &3.92\err{0.10}{0.20} &2.87\err{0.28}{0.10}	 &3.12\err{0.09}{0.12}	 &3.72\err{0.31}{0.27}	 &5.90\err{0.27}{0.35}	 &5.28\err{0.06}{0.07}	 &4.70\err{0.04}{0.20}	 &3.85\err{0.15}{0.04}        \\
Redshift\tablenotemark{b} ($10^{-3}$)          &$-$2.05	 &$-$1.94		 &$-$2.81	 &2.82	 &$-$3.17	 &$-$2.11  &$-$3.22	 &$-$3.30	 &$-$3.22	 &$-$1.94     \\
Norm\tablenotemark{f} ($10^{-3}$ cm$^{-5}$) &6.57			 &8.71			 &10.02			 &8.27			 &6.44			 &3.04			 &3.88			 &6.39			 &10.94		 &9.62                \\
$\chi^2$/dof	                  &1.18			 &1.25			 &1.19			 &1.19			 &1.19			 &1.12			 &1.12			 &1.17			 &1.24			 &1.29                     \\
\enddata
\end{deluxetable*}

\begin{deluxetable*}{lllllllllll} \rotate
\tablenum{2}
\tabletypesize{\footnotesize}
\tablecolumns{11}
\tablewidth{0pt} 
\tablecaption{Best-fit Model Parameters\label{table:fits}} 
\tablehead{ \colhead{Parameter} & \colhead{Region 21}& \colhead{Region 22}& \colhead{Region 23}& \colhead{Region 24}& \colhead{Region 25}& \colhead{Region 26}& \colhead{Region 27}& \colhead{Region 28}& \colhead{Region 29}& \colhead{Region 30}}
\startdata 
$N_{\rm H}$ (10$^{22}$ cm$^{-2}$)      &8.04\err{0.02}{0.01}	 &8.03\err{0.02}{0.01}	 &7.77\err{0.04}{0.02}	 &7.51\err{0.06}{0.09}	 &7.74\err{0.02}{0.02}	 &7.83\err{0.03}{0.03}&7.75\err{0.03}{0.03}& 8.30\err{0.08}{0.05}	 &8.08\err{0.01}{0.04}	 &7.97\err{0.03}{0.03}    \\
\cutinhead{ISM \texttt{vapec}\tablenotemark{a} component} 
$kT_{\rm e}$ (keV) 	                  &0.173\err{0.001}{0.001}&0.178\err{0.001}{0.001} &0.1900\err{0.001}{0.001} &0.181\err{0.002}{0.002}&0.173\err{0.001}{0.001}&0.174\err{0.001}{0.001} &0.177\err{0.001}{0.002}&0.166\err{0.003}{0.002}&0.178\err{0.001}{0.001}&0.183\err{0.001}{0.001}\\
Redshift\tablenotemark{b} ($10^{-3}$)                &$-$5.37			 &$-$2.73			 &3.91			 &6.54			 &5.10			 &3.70			 &3.18			 & 5.74			 &0.13			 &0.01    \\
Norm\tablenotemark{b} 	(cm$^{-5}$)             &12.29			 &8.41			 &3.30			 &2.65			 &5.03			 &8.15			 &6.02			 & 12.63 			 &7.41		 &4.87    \\
\cutinhead{Ejecta 1 \texttt{vvapec} or \texttt{vvrnei}\tablenotemark{c} component\tablenotemark{d}}
$kT_{\rm e}$ (keV) & 0.51\err{0.01}{0.02} & 0.51\err{0.01}{0.01} & 0.46\err{0.01}{0.01}&0.39\err{0.03}{0.03}&0.55\err{0.02}{0.01}&0.54\err{0.01}{0.04}&0.55\err{0.01}{0.03}& 0.45\err{0.01}{0.02}&0.50\err{0.01}{0.01}&0.51\err{0.01}{0.01}   \\
$\tau_{1}$ ($10^{11}$ cm$^{-3}$ s) &9.45\err{0.09}{0.21}	 &6.76\err{0.09}{0.09}	 &7.06\err{0.10}{0.08}	 &8.05\err{0.67}{0.59}	 & ---			 & ---			 & --- 			 & 10.57\err{0.45}{0.18} &6.84\err{0.07}{0.11}	 &5.50\err{0.08}{0.06}  \\
Redshift\tablenotemark{e} ($10^{-3}$)          &0.14				 &$-$4.26			 &$-$4.24			 &2.77			 &$-$5.24			 &$-$2.47			 &$-$2.53 			 & $-$0.07			 &$-$4.31			 &$-$4.31   \\
Norm ($10^{-3}$ cm$^{-5}$)    &8.04\err{0.11}{0.37}	 &9.84\err{0.15}{0.09}	 &8.09\err{0.41}{0.56}	 &4.57\err{0.89}{0.50}	 &2.98\err{0.20}{0.22}	 &5.15\err{0.90}{0.47}	 &5.19\err{0.35}{0.14} 	 & 10.63\err{0.96}{-0.23}&11.57\err{0.80}{0.94}	 &10.83\err{0.07}{0.07}   \\
\cutinhead{Ejecta 2 \texttt{vvrnei}\tablenotemark{e} component}
$kT_{\rm e}$ (keV)                  &1.29\err{0.01}{0.01} &1.21\err{0.01}{0.01}&1.04\err{0.01}{0.01}&0.87\err{0.04}{0.09}&1.50\err{0.02}{0.02}&1.52\err{0.01}{0.02}&1.45\err{0.01}{0.02}& 1.30\err{0.02}{0.02}&1.33\err{0.03}{0.04}&1.36\err{0.01}{0.01}   \\
$kT_{\rm init}$ (keV)         &3.74\err{0.02}{0.08}	 &3.619\err{0.017}{0.027}&3.562\err{0.085}{0.029}&3.81\err{0.034}{0.041} &4.69\err{0.11}{0.26}	 &4.59\err{0.07}{0.07}	 &4.26\err{0.07}{0.15} 	 & 4.01\err{0.04}{0.18}  &3.78\err{0.08}{0.11}	 &3.84\err{0.04}{0.04}   \\
Mg	                        &4.06\err{0.45}{0.42}	 &5.16\err{0.51}{0.48}	 &3.76\err{0.48}{0.35}	 &5.86\err{1.02}{0.98}	 &8.78\err{0.54}{0.51}	 &10.2\err{0.7}{0.6}	 &5.46\err{0.40}{0.53} 	 & 5.09\err{0.62}{0.68}  &2.55\err{0.47}{0.31}	 &3.65\err{0.75}{0.37}   \\
Si		                  &13.1\err{0.13}{0.30}	 &12.0\err{0.2}{0.3}	 &12.1\err{0.5}{0.2}	 &12.5\err{0.6}{1.1}	 &11.7\err{0.1}{0.3}	 &13.8\err{0.2}{0.2}	 &11.54\err{0.16}{0.17}  & 11.5\err{0.7}{0.4}    &8.14\err{0.35}{0.18}	 &8.74\err{0.41}{0.41}    \\
S 	                        &16.6\err{0.22}{0.30}	 &16.7\err{0.2}{0.3}	 &17.4\err{0.2}{0.3}	 &17.3\err{0.7}{1.5}	 &13.3\err{0.2}{0.2}	 &15.6\err{0.2}{0.2}	 &13.9\err{0.14}{0.15} 	 & 13.7\err{0.6}{0.3}    &11.4\err{0.4}{0.4}	 &12.6\err{0.5}{0.2}    \\
Ar	                        &18.2\err{0.44}{0.34}	 &18.0\err{0.4}{0.4}	 &19.8\err{0.4}{0.6}	 &19.8\err{1.3}{1.9}	 &13.7\err{0.2}{0.2}	 &16.9\err{0.2}{0.6}	 &15.9\err{0.36}{0.37} 	 & 15.3\err{0.4}{0.5}    &11.9\err{0.4}{0.3}	 &13.8\err{0.4}{0.4}    \\
Ca	                        &18.9\err{0.38}{0.34}	 &18.5\err{0.5}{0.4}	 &18.9\err{0.4}{0.5}	 &18.9\err{1.3}{1.0}	 &14.8\err{0.4}{0.5}	 &17.9\err{0.3}{0.5}	 &15.6\err{0.34}{0.35} 	 & 16.0\err{0.4}{0.5}    &12.4\err{0.4}{0.3}	 &13.5\err{0.4}{0.3}   \\
Cr	                        &33.2\err{2.43}{2.56}	 &35.2\err{2.9}{3.2}	 &43.9\err{4.5}{4.8}	 &50.3\err{43.0}{31.2}	 &25.3\err{3.7}{2.6}	 &28.0\err{2.1}{3.1}	 &21.1\err{2.3}{2.5} 	 & 21.2\err{2.9}{2.9}	 &15.8\err{2.4}{2.2}	 &23.7\err{2.9}{2.9}    \\
Mn	                        &60.9\err{4.55}{6.88}	 &51.3\err{6.8}{9.9}	 &49.1\err{9.4}{14.1}	 &47.7\err{38.1}{16.5}	 &50.2\err{6.7}{7.4}	 &53.4\err{6.1}{5.9}	 &32.3\err{5.4}{5.5} 	 & 22.9\err{6.8}{6.8}    &33.0\err{4.2}{6.9}	 &23.4\err{4.3}{4.3}     \\
Fe	                        &25.1\err{0.16}{0.55}	 &16.0\err{0.2}{0.2}	 &12.8\err{0.2}{0.2}	 &13.0\err{1.2}{1.4}	 &18.6\err{0.5}{0.2}	 &24.8\err{0.2}{0.2}	 &20.9\err{0.4}{0.4} 	 & 18.3\err{0.3}{0.2}    &12.3\err{0.5}{0.2}	 &9.52\err{0.14}{0.13}     \\
Ni	                        &58.0\err{3.57}{2.83}	 &28.2\err{4.1}{3.7}	 &14.4\err{5.1}{5.1}	 &$<$3.37			 &49.6\err{4.0}{3.5}	 &76.8\err{2.6}{4.3}	 &62.7\err{2.8}{3.6} 	 & 52.5\err{4.9}{7.7}    &33.2\err{ 4.7}{2.8}	 &12.3\err{3.7}{3.7}     \\
$\tau_{2}$ ($10^{11}$ cm$^{-3}$ s) &3.13\err{0.03}{0.09}	 &2.53\err{0.09}{0.09}	 &2.84\err{0.04}{0.03}	 &4.02\err{0.32}{0.56}	 &6.43\err{2.2}{1.0}	 &6.34\err{0.10}{0.05}	 &5.35\err{0.19}{0.21} 	 & 4.27\err{0.07}{0.40}	 &3.17\err{0.05}{0.11}	 &3.03\err{0.22}{0.22}      \\
Redshift\tablenotemark{b} ($10^{-3}$)          &$-$2.81			 &$-$2.81			 &$-$2.82			 &$-$2.35			 &$-$3.22			 &$-$3.23			 &$-$3.24			 &$-$3.24			 &$-$2.02			 &$-$1.94     \\
Norm\tablenotemark{f} ($10^{-3}$ cm$^{-5}$) &7.64				 &5.96			 &4.84			 &3.53			 &4.00			 &5.96			 &6.14 			 & 6.10	          	 &5.43			 &4.05     \\
$\chi^2$/dof	                  &1.24			 &1.16			 &1.19			 &1.09			 &1.14			 &1.19			 &1.24			 & 1.22             	 &1.10			 &1.10     \\
\enddata
\end{deluxetable*}

\begin{deluxetable*}{lllllllllll} \rotate
\tablenum{2}
\tabletypesize{\footnotesize}
\tablecolumns{11}
\tablewidth{0pt} 
\tablecaption{Best-fit Model Parameters\label{table:fits}} 
\tablehead{ \colhead{Parameter} & \colhead{Region 31}& \colhead{Region 32}& \colhead{Region 33}& \colhead{Region 34}& \colhead{Region 35}& \colhead{Region 36}& \colhead{Region 37}& \colhead{Region 38}& \colhead{Region 39}& \colhead{Region 40}}
\startdata 
$N_{\rm H}$ (10$^{22}$ cm$^{-2}$)      &7.76\err{0.09}{0.07}	 &7.19\err{0.06}{0.04} &7.38\err{0.02}{0.03}	 	 &7.27\err{0.02}{0.06}	 &7.50\err{0.05}{0.07}	 &7.68\err{0.07}{0.06}	 &8.15\err{0.08}{0.09}	 &8.06\err{0.09}{0.05}	 &7.48\err{0.12}{0.10}	 &7.18\err{0.04}{0.02}     \\
\cutinhead{ISM \texttt{vapec}\tablenotemark{a} component} 
$kT_{\rm e}$ (keV) 	   &0.175\err{0.002}{0.001}&0.177\err{0.002}{0.001}&0.1906\err{0.001}{0.002}&0.189\err{0.002}{0.002}&0.189\err{0.002}{0.002}&0.189\err{0.003}{0.003}&0.177\err{0.001}{0.002}&0.174\err{0.001}{0.002}&0.178\err{0.003}{0.002}&0.195\err{0.001}{0.005} \\
Redshift\tablenotemark{b} ($10^{-3}$)                &6.72				 &5.52			 &$-$5.66				 &2.56			 &$-$0.72			 &$-$5.32			 &$-$4.94			 &6.89			 &6.64			 &$-$0.74	       \\
Norm\tablenotemark{d} 	(cm$^{-5}$)             &4.97			 &2.32			 &2.84				 &1.87			 &1.83			 &1.93			 &4.78			 &4.40			 &1.92			 &1.21	     \\
\cutinhead{Ejecta 1 \texttt{vvapec} or \texttt{vvrnei}\tablenotemark{c} component\tablenotemark{d}}
$kT_{\rm e}$ (keV)        			&0.48\err{0.02}{0.03} &0.47\err{0.02}{0.02}&0.76\err{0.02}{0.01}	 &0.60\err{0.01}{0.01}&0.57\err{0.02}{0.05}&0.43\err{0.03}{0.01}&0.47\err{0.02}{0.02}&0.49\err{0.03}{0.01}&0.46\err{0.03}{0.03}&0.75\err{0.01}{0.02}     \\
$\tau_{1}$ ($10^{11}$ cm$^{-3}$ s) &8.06\err{0.66}{0.41}	 &8.57\err{0.26}{0.20}	 & ---			 & ---			 & ---		 &10.2\err{0.44}{0.54}	 &8.11\err{0.33}{0.44}	 &7.66\err{0.52}{0.27}	 &8.42\err{0.64}{0.58}	 & ---       \\
Redshift\tablenotemark{b} ($10^{-3}$)          &0.15				 &0.16			 &$-$1.94				 &$-$2.12			 &$-$2.13			 &$-$0.64			 &$-$1.34			 &0.06			 &$-$0.14			 &$-$2.47	   \\
Norm ($10^{-3}$ cm$^{-5}$)    &6.83\err{0.56}{0.67}	 &5.73\err{0.37}{0.37}	 &3.91\err{0.10}{0.12}	 	 &5.59\err{0.55}{0.79}	 &3.34\err{0.34}{0.28}	 &9.06\err{0.80}{1.01}	 &9.90\err{0.79}{0.77}	 &5.06\err{0.72}{0.54}	 &3.55\err{0.54}{0.41}	 &2.04\err{0.07}{0.05}        \\
\cutinhead{Ejecta 2 \texttt{vvrnei}\tablenotemark{e} component}
$kT_{\rm e}$ (keV)                    &0.97\err{0.02}{0.03}&0.90\err{0.02}{0.02}&1.53\err{0.02}{0.01}	 &1.39\err{0.02}{0.07}&1.00\err{0.01}{0.01}&1.05\err{0.02}{0.05}&1.07\err{0.03}{0.04}&0.99\err{0.04}{0.03}&0.90\err{0.05}{0.04}&1.54\err{0.03}{0.02}      \\
$kT_{\rm init}$ (keV)         &3.25\err{0.08}{0.15}	 &3.38\err{0.09}{0.23}	 &4.06\err{0.07}{0.08}	 	 &4.42\err{0.08}{0.25}	 &3.61\err{0.10}{0.10}	 &3.32\err{0.32}{0.21}	 &3.54\err{0.20}{0.22}	 &3.38\err{0.21}{0.11}	 &3.58\err{0.28}{0.25}	 &4.07\err{0.16}{0.09}     \\
Mg	                        &2.85\err{0.69}{0.71}	 &3.28\err{0.45}{0.35}	 &7.04\err{0.44}{0.38}	 	 &5.62\err{0.36}{0.36}	 &5.78\err{0.15}{0.54}	 &4.15\err{0.38}{0.45}	 &3.94\err{0.62}{0.74}	 &5.15\err{0.96}{0.57}	 &5.04\err{0.93}{0.93}	 &6.53\err{0.47}{0.66}         \\
Si		                  &11.3\err{0.7}{0.5}	 &10.3\err{0.5}{0.6}	 &8.91\err{0.20}{0.12}	 	 &9.36\err{0.15}{0.16}	 &12.1\err{0.6}{0.3}	 &9.86\err{0.26}{0.25}	 &8.21\err{0.64}{0.71}	 &12.0\err{0.9}{1.3}	 &14.1\err{1.2}{1.4}	 &9.61\err{0.18}{0.24}           \\
S 	                        &15.9\err{1.0}{0.7}	 &15.7\err{1.6}{0.3}	 &12.0\err{0.2}{0.2}		 &12.2\err{0.2}{0.1}	 &15.5\err{0.6}{0.6}	 &13.4\err{0.88}{0.94}	 &11.5\err{0.9}{0.9}	 &17.2\err{0.6}{0.8}	 &21.5\err{2.0}{1.0}	 &12.9\err{0.2}{0.3}           \\
Ar	                        &18.6\err{1.2}{0.9}	 &19.4\err{0.7}{1.1}	 &13.8\err{0.3}{0.4}		 &15.4\err{0.4}{0.3}	 &18.2\err{0.7}{0.8}	 &16.2\err{0.6}{1.3}	 &13.3\err{1.2}{1.4}	 &20.0\err{1.5}{1.8}	 &25.4\err{2.5}{2.8}	 &15.1\err{0.5}{0.5}          \\
Ca	                        &18.2\err{1.3}{0.9}	 &17.5\err{1.5}{0.9}	 &13.6\err{0.3}{0.4}		 &14.9\err{0.3}{0.4}	 &15.6\err{0.7}{0.6}	 &16.8\err{0.7}{0.7}	 &14.3\err{1.1}{1.4}	 &19.5\err{1.7}{1.6}	 &24.9\err{2.5}{2.5}	 &13.9\err{0.5}{0.4}          \\
Cr	                        &30.8\err{5.6}{4.1}	 &27.4\err{6.4}{6.1}	 &18.7\err{2.0}{2.1}		 &22.7\err{ 2.5}{2.6}	 &33.4\err{5.3}{6.0}	 &17.8\err{4.4}{4.3}	 &14.1\err{4.5}{3.9}	 &61.7\err{10.8}{17.6}	 &88.6\err{52.3}{35.7}	 &14.7\err{2.5}{2.6}          \\
Mn	                        &$<$150 			 &49.3\err{9.1}{15.1}	 &39.5\err{5.0}{4.9}		 &36.5\err{6.1}{5.7}	 &35.4\err{14.7}{15.1}	 &29.5\err{13.5}{14.2}	 &27.2\err{9.4}{11.2}	 &46.7\err{15.2}{14.1}	 &40.8\err{17.7}{18.1}	 &27.8\err{5.4}{5.4}           \\
Fe	                        &11.5\err{1.0}{0.7}	 &9.83\err{1.65}{0.42}	 &14.5\err{0.2}{0.2}		 &16.2\err{0.2}{0.4}	 &14.0\err{0.7}{0.5}	 &11.2\err{0.7}{1.1}	 &7.88\err{0.87}{0.92}	 &11.8\err{1.4}{1.3}	 &13.5\err{1.5}{1.8}	 &14.9\err{0.4}{0.7}          \\
Ni	                        &10.8\err{8.5}{6.5}	 &$<4.75$			 &29.4\err{3.8}{2.4}		 &50.6\err{3.0}{3.1}	 &36.4\err{12.4}{10.8}	 &41.1\err{5.6}{16.3}	 &$<$7.42 			 &19.4\err{13.1}{10.9}	 &17.2\err{15.3}{7.7}	 &23.5\err{3.1}{3.2}           \\
$\tau_{2}$ ($10^{11}$ cm$^{-3}$ s) &2.39\err{0.25}{0.29}	 &2.41\err{0.29}{0.14}	 &4.19\err{0.20}{0.09}	 	 &4.35\err{0.18}{0.18}	 &2.89\err{0.19}{0.23}	 &2.46\err{0.33}{0.42}	 &2.33\err{0.28}{0.30}	 &2.44\err{0.52}{0.27}	 &2.46\err{0.34}{0.37}	 &3.73\err{0.25}{0.26}        \\
Redshift\tablenotemark{b} ($10^{-3}$)          &$-$2.15			 &$-$2.30			 &$-$3.20				 &$-$3.20			 &$-$2.82			 &$-$3.16			 &$-$3.33			 &$-$2.41			 &$-$2.38			 &$-$2.81	     \\
Norm\tablenotemark{f} ($10^{-3}$ cm$^{-5}$) &5.69			 &3.84			 &5.75				 &5.92			 &4.03			 &3.62			 &4.13			 &3.43			 &2.22			 &2.98	               \\
$\chi^2$/dof	                  &1.11			 &1.13			 &1.17				 &1.18			 &1.13			 &1.12			 &1.11			 &1.06			 &1.05			 &1.10	               \\
\enddata
\end{deluxetable*}

\begin{deluxetable*}{lllllll} \rotate
\tablenum{2}
\tabletypesize{\footnotesize}
\tablecolumns{7}
\tablecaption{Best-fit Model Parameters\label{table:fits}} 
\tablehead{\colhead{Parameter} & \colhead{Region 41}& \colhead{Region 42}& \colhead{Region 43}& \colhead{Region 44}& \colhead{Region 45}& \colhead{Region 46}}
\startdata 
$N_{\rm H}$ (10$^{22}$ cm$^{-2}$)      & 7.36\err{0.03}{0.01} & 7.07\err{0.03}{0.02} & 7.13\err{0.03}{0.03} & 7.47\err{0.10}{0.10} & 7.83\err{0.03}{0.08} & 7.42\err{0.17}{0.12}     \\
\cutinhead{ISM \texttt{vapec}\tablenotemark{a} component} 
$kT_{\rm e}$ (keV) 	                  & 0.188\err{0.001}{0.001} & 0.191\err{0.002}{0.001} & 0.188\err{0.001}{0.002} & 0.185\err{0.001}{0.001} & 0.177\err{0.001}{0.002} & 0.177\err{0.002}{0.002}  \\
Redshift\tablenotemark{b} ($10^{-3}$)                & $-$0.73 & $-$0.74 & 0.22 & 7.06 & 7.52 & $-$4.80      \\
Norm\tablenotemark{b} 	(cm$^{-5}$)             & 2.38 & 1.13 & 1.17 & 1.41 & 1.93 & 0.77       \\
\cutinhead{Ejecta 1 \texttt{vvapec} or \texttt{vvrnei}\tablenotemark{c} component\tablenotemark{d}}
$kT_{\rm e}$ (keV)        			& 0.73\err{0.01}{0.02} & 0.52\err{0.03}{0.02} & 0.54\err{0.02}{0.03} & 0.56\err{0.04}{0.02} & 0.49\err{0.03}{0.02}  & 0.43\err{0.02}{0.02} \\
$\tau_{1}$ ($10^{11}$ cm$^{-3}$ s) & --- & 14.44\err{0.43}{0.32} & 12.22\err{0.37}{0.41} & 9.18\err{0.18}{0.40} & 8.03\err{0.50}{0.15}  & 7.40\err{0.34}{0.13} \\
Redshift\tablenotemark{b} ($10^{-3}$)          & $-$1.92 & $-$3.03 & $-$2.12 & $-$2.66 & $-$1.92 & 0.14 \\
Norm ($10^{-3}$ cm$^{-5}$)    & 3.03\err{0.07}{0.04} & 7.70\err{0.31}{0.20} & 6.54\err{0.26}{0.18} & 4.53\err{0.77}{0.77} & 5.34\err{0.70}{0.27}  & 4.71\err{0.21}{0.50}        \\
\cutinhead{Ejecta 2 \texttt{vvrnei}\tablenotemark{e} component}
$kT_{\rm e}$ (keV)                    & 1.44\err{0.01}{0.01} & 1.02\err{0.03}{0.01} & 0.92\err{0.03}{0.03} & 1.00\err{0.04}{0.04} & 0.96\err{0.06}{0.02} & 0.97\err{0.03}{0.04}     \\
$kT_{\rm init}$ (keV)         & 3.84\err{0.08}{0.04} & 3.87\err{0.16}{0.09} & 3.73\err{0.07}{0.06} & 2.95\err{0.26}{0.18} & 2.72\err{0.15}{0.20}  & 3.28\err{0.12}{0.12}   \\
Mg	                        & 6.11\err{0.49}{0.43} & 2.58\err{0.28}{0.23} & 1.86\err{0.25}{0.24} & 3.41\err{0.48}{0.44} & 2.81\err{0.50}{0.54}  & 2.50\err{0.53}{0.38}        \\
Si		                  & 11.2\err{0.2}{0.2} & 6.01\err{0.15}{0.16} & 4.47\err{0.13}{0.14} & 5.98\err{0.70}{0.70} & 7.25\err{0.32}{0.24}  & 7.70\err{0.28}{0.36}         \\
S 	                        & 14.7\err{0.2}{0.3} & 8.00\err{0.17}{0.11} & 6.28\err{0.16}{0.13} & 8.50\err{0.29}{0.65} & 10.7\err{0.28}{0.28}  & 12.2\err{0.9}{0.4}          \\
Ar	                        & 17.5\err{0.4}{0.5} & 9.58\err{0.39}{0.45} & 7.04\err{0.36}{0.35} & 9.89\err{0.60}{0.52} & 12.3\err{1.6}{0.8}  & 13.8\err{0.9}{0.9}         \\
Ca	                        & 14.9\err{0.4}{0.5} & 9.39\err{0.44}{0.45} & 7.51\err{0.39}{0.39} & 11.1\err{0.71}{0.65} & 15.1\err{0.8}{0.8}  & 15.9\err{1.0}{1.2}         \\
Cr	                        & 15.3\err{2.8}{2.8} & 14.2\err{3.7}{3.2} & 12.6\err{3.4}{3.5} & 25.2\err{6.5}{6.1} & 37.8\err{7.8}{7.6}  & 33.5\err{8.5}{9.0}        \\
Mn	                        & 22.2\err{5.8}{5.7} & $<$8.82 & $<$8.20 & 18.8\err{13.9}{13.4} & 32.7\err{17.5}{16.8}  & $<$22.9         \\
Fe	                        & 15.7\err{0.2}{0.2} & 7.53\err{0.21}{0.62} & 5.44\err{0.65}{0.23} & 7.74\err{1.21}{0.33} & 8.00\err{0.30}{0.49}  & 8.49\err{0.47}{1.90}        \\
Ni	                        & 28.8\err{4.1}{2.8} & $<$3.34 & $<$2.47 & $<$3.56 & $<$5.74 &  $<$16.5   \\
$\tau_{2}$ cm$^{-3}$ s) & 2.84\err{0.10}{0.06} & 3.03\err{0.23}{0.13} & 2.88\err{0.10}{0.08} & 2.12\err{0.76}{0.18} & 1.40\err{0.26}{0.59}  &  2.23\err{0.20}{0.39}       \\
Redshift\tablenotemark{b} ($10^{-3}$)          & $-$2.81 & $-$2.90 & $-$3.33 & $-$2.81 & $-$2.80 & $-$2.79     \\
Norm\tablenotemark{f} ($10^{-3}$ cm$^{-5}$) & 3.48 & 3.94 & 4.13 & 2.36 & 2.14 & 1.79 \\
$\chi^2$/dof	                  & 1.18 & 1.10 & 1.13 & 1.15 & 1.14 & 1.02 \\
\enddata
\tablenotetext{a}{All abundances frozen to 1, except for Mg which was frozen to 0.3 to match the analysis of \cite{sun20}}
\tablenotetext{b}{Frozen after initial fit to better constrain the other parameters}
\tablenotetext{c}{$kT_{\rm init}$ tied to the initial temperature of the 2nd ejecta component}
\tablenotetext{d}{Element abundances linked to the Ejecta 2 abundances}
\tablenotetext{e}{Unspecified abundances frozen to 1}
\tablenotetext{f}{Unconstrained and thus frozen after initial fit}
\end{deluxetable*}

\end{appendix}

\nocite{*}
\bibliographystyle{apj}
\bibliography{w49b}

\end{document}